\input harvmac
\input epsf.tex
\ifx\answ\bigans

\else

\fi

\def\caption#1{
        \centerline{\vbox{\baselineskip=12pt
        \vskip.15in\hsize=3.8in\noindent{#1}\vskip.1in }}}


\def\ol{\overline}

\def\O{{\cal O}}

\def\d{{\rm d}}
\def\im{{\rm i}}
\def\q{{\rm q}}

\def\kslash{k\hskip-0.5em /}

\def\OMIT#1{}
\def\frac#1#2{{#1\over#2}}
\def\eps{{\epsilon}}


\def\pd#1{{\partial \over \partial #1}}
\def\alpi{{\alpha \over 3 \pi}}
\def\lom{$\Lambda_{QCD}\over m_b$}

\def\ll{\frac12 \ln 4\pi}
\def\dz{\delta Z}
\def\D{\Delta_0}
\def\ub{\overline u_b(p)}
\def\u{u_b(p)}
\def\NV{N^{(V)} }
\def\NB{N^{(B)} }
\def\lint{\int{\d^D\,l\over(2\pi)^D}}
\def\gprop{[(l-p)^2 - \lambda^2+ \im \epsilon]}
\def\bprop{[l^2 - m_b^2 + \im \epsilon]}
\def\cprop{[(l-q)^2 - m_j^2 + \im \epsilon]}
\def\ga{\gamma}
\def\al{\alpha}
\def\be{\beta}
\def\la{\lambda}
\def\qslash{q\hskip-0.5em /}
\def\pslash{p\hskip-0.5em /}
\def\lslash{l\hskip-0.5em /}
\def\vslash{v\hskip-0.5em /}
\def\vq{\hbox{v\kern+0.1em $\cdot$ \kern-.3em q}}
\def\ba{{\bf a}}

\def\vk{\hbox{v\kern+0.1em $\cdot$ \kern-.3em k}}
\def\vx{\hbox{v\kern+0.1em $\cdot$ \kern-.3em x}}
\def\qs{\hbox{s\kern+0.1em $\cdot$ \kern-.3em q}}
\def\qx{\hbox{q\kern+0.1em $\cdot$ \kern-.3em x}}

\def\tpid{{2 \pi}^{D-4}}


\def\pl#1#2#3{{Phys.~Lett.}~{#1}B (#2) #3}

\def\physrev#1#2#3{{Phys.~Rev.}~{#1} (#2) #3}

\def\bar#1{\overline{#1}}

\def\bra#1{\left\langle #1\right|}
\def\ket#1{\left| #1\right\rangle}
\def\abs#1{\left| #1\right|}

\def\Tr{\mathop{\rm Tr}}

\def\CA{{\cal A}} \def\CB{{\cal B}}

  \def\CO{{\cal O}}

\def\[{\left[}
\def\]{\right]}
\def\({\left(}
\def\){\right)}

\def\[{\left[}
\def\]{\right]}

\def\sa{S_{,a}}
\def\sb{S_{,b}}
\def\sc{S_{,c}}
\def\sd{S_{,d}}
\def\se{S_{,e}}
\def\sf{S_{,f}}
\def\saa{S_{,a,a}}
\def\sab{S_{,a,b}}
\def\sac{S_{,a,c}}
\def\sad{S_{,a,d}}
\def\sae{S_{,a,e}}
\def\saf{S_{,a,f}}
\def\scc{S_{,c,c}}
\def\scd{S_{,c,d}}
\def\sce{S_{,c,e}}
\def\sde{S_{,d,e}}
\def\tscoef{{\alpha \over 3 \pi}}
\def\ovrep{\frac1\epsilon-\frac\gamma2+\frac12\ln(4\pi)}
\def\en{E_\nu}
\def\et{E_\tau}
\def\mts{m_\tau^2}
\def\kleng{\sqrt{\et^2-\mts}}
\def\dcero{(-4+3 \ln m_b^2 -2 \ln \lambda^2)}
\def\dt{(m_b^2+q^2-2 m_b \vq-m_j^2)}

\def\ms{\CB[\vq,m_j^2]}

\Title{\vbox{\hbox{UCSD/PTH 94-22}}}
{Strong Corrections to
Inclusive $B\rightarrow X\tau\bar\nu_{\tau}$  Decays}
\vskip .5cm
\centerline{\bf{C. Glenn Boyd,  F. Javier Vegas}}
{\centerline{Dept. of Physics 0319, University of
California at San Diego}
\centerline{9500 Gilman Dr, La Jolla, CA 92093}}
\vskip .2cm
\centerline{\bf{Zachary Guralnik}}
\centerline{Physics Department}
\centerline{Princeton University}
\centerline{Princeton, NJ 08544}
\vskip .2cm
\centerline{and}
\centerline{\bf{Martin Schmaltz}}
\centerline{Institute for Nuclear Theory, NK-12}
\centerline{University of Washington}
\centerline{Seattle, WA  98195}

\vskip .5cm
\centerline{\bf{Abstract}}
\vskip .3in
We calculate the $\alpha_s$ corrections to the
form factors which parameterize the hadronic tensor relevant for
inclusive semileptonic $B \rightarrow X \tau\bar\nu_{\tau}$ and
$\Lambda_b \rightarrow X \tau\bar\nu_{\tau}$ decays.
We apply  our results  to  the double differential decay rates for the
decays of $B$ mesons and polarized $\Lambda_b$ baryons to polarized
$\tau$ leptons, presenting them
in terms of one-dimensional integrals. Formulas appropriate for
insertion into a Monte-Carlo simulation are presented.

\vskip .3in
\Date{Dec. 1994}
\newsec{Introduction}
There has been much recent interest in inclusive semileptonic
$B$ decays as a means of measuring the CKM parameters $V_{ub}$ and
$V_{cb}$.  These decays are very difficult to calculate due to
quark confinement and the breakdown of perturbation theory.
However it is possible to improve the behavior of the QCD perturbation
series by considering a ``smeared'' decay rate, in which
one integrates in a certain way over the lepton momenta.
The infrared singularities responsible for
the breakdown of pertubation
theory are tamed
by integrating appropriately across perturbative thresholds.
The idea of smearing to improve the behavior of the
QCD pertubation series in the physical region has been used for
example to examine the
 inclusive process $e^+e^-\rightarrow$
hadrons~\ref\pqw{
E.C. Poggio,  H.R. Quinn,
and S. Weinberg, \physrev{D13}{1976}{1958}.}\
For this process one computes averages of the QCD
contribution to the vacuum polarization.
This smearing method can also be used
to describe inclusive semileptonic heavy meson
decays $B\rightarrow Xl\bar{\nu}$ ~\ref\cgg{J. Chay,  H. Georgi, and
B. Grinstein, Phys. Lett. B 247, (1990) 399.}.
In this case there are additional
non-perturbative complications, since one must compute the expectation
value of quark currents in a $B$ meson rather than the vacuum.
This difficulty is controlled by means of a heavy quark expansion.
In this paper, we compute perturbative QCD corrections to
semi-inclusive semileptonic heavy meson decays involving tau
leptons, in the framework of the above smearing method. We
will consider the leading term in the heavy quark expansion, so that
in practice no non-perturbative matrix elements will be needed.

The semileptonic $B$ meson decays are caused by the
Weak Hamiltonian density
\eqn\hweke{H_W = -V_{jb}{4G_F\over\sqrt{2}}
\bar q_j\gamma^{\mu}P_Lb\bar l \gamma_{\mu}P_L\nu,}
where $P_L$ is the chiral projector ${1\over2}(1-\gamma_5)$.
The index $j$ denotes either a charm or an up type quark.
The inclusive decay rate is given by
\eqn\ink{{d \Gamma \over dq^2dE_{\nu}dE_{\tau}}=
{\abs {V_{j b}}^2
G_f^2\over 2\pi ^3} L_{\mu\nu}W^{\mu\nu}.}
$L_{\mu\nu}$ is an easily computed matrix element of
leptonic weak currents
\eqn\lmunu{L_{\mu \nu} = Tr[( \pslash_l+m_l) \gamma_\mu (\pslash_\nu)
         \gamma_\nu P_L],}
and $W_{\mu\nu}$ is the hadronic matrix element
\eqn\had{{W_{\mu\nu}=(2\pi)^3\sum_X\delta^4(p_B -q-p_X)
         \bra{B(v,s)}J_j^{\mu}\ket{X}\bra{X}
         J_j^{\nu}\ket{B(v,s)}},}
with $J_j^{\mu}$  the weak quark current $\bar u_j\gamma^{\mu}P_Lb$
and $q=p_l+p_\nu$  the total lepton momentum.
Using the optical theorem,
$W^{\mu\nu}$ may be expressed as the discontinuity across the
cut
in the time ordered
product  of weak currents in a B meson:
\eqn\WT{W^{\mu\nu}=-{1\over\pi}Im T^{\mu\nu},}
where \eqn\tord{T^{\mu\nu} = -i\int d^4x \exp (i\qx)
\bra{B}T[J^{\mu^{\dagger}}(x)J^{\nu}(0)]\ket{B}.}
The B quark matrix element may be computed
using a heavy quark expansion in ${\Lambda_{QCD} \over m_b}$.

In the approach of ref~\cgg\ ,
there are two steps to this expansion.
The first consists of
an operator product expansion in which one expands about
an ``on-shell'' value for the heavy quark momentum:
$p=m_bv+k$ where $m_b$ is the mass parameter
of the heavy quark effective
lagrangian, $v$ is the four velocity
of the $B$ meson, and $k$ is a small residual off-shell momentum.
The Wilson coefficients are written as a pertubation series in
$\alpha_s$.  The second step is a heavy quark expansion of the
expectation values of the local operators in a B meson.
The Wilson coefficients are singular at the perturbative
threshold, at which the invariant mass of the
hadronic decay products is $m_j$.  However, by appropriately
smearing the decay rate over lepton energies
the operator product expansion can be made well behaved simultaneously
with the
pertubative QCD expansion. This will be discussed in more detail in
section 2.  The lowest order term in the
${\Lambda_{QCD}\over m_b}$ expansion is simply the parton model result
for the appropriately spin averaged decay of a $b$ quark.

It has been shown that the $\Lambda_{QCD}\over m_b$ corrections
vanish~\cgg,
which lends considerable credence to smeared parton model calculations.
The order
$({\Lambda_{QCD}\over m_b})^2$ corrections have been computed at tree
level~\ref\bksv{A.V. Manohar and M.B. Wise, \physrev{D49}{1994}{1310}\semi
I.I. Bigi,  M. Shifman, N.G. Uraltsev and A.I. Vainshtein,
Phys. Rev. Lett. 71, (1993) 496\semi
B. Blok, L. Koyrakh, M. Shifman and A.I. Vainshtein, Phys. Rev. D49,
3356 (1994)\semi
T. Mannel, Nucl. Phys. B413, 396 (1994).}, and the resulting parametrization
has been applied to the differential decay rate for $\ol B \to X \tau \ol \nu$
with polarized and unpolarized tau leptons\nref\flz{
A.F. Falk, Z. Ligeti, and M. Neubert, \pl{326}{1994}{145}.}.
To order $\alpha_s$ the parton model result for semileptonic
decays involving an electron have been calculated\nref\pwk{
N. Cabibbo,  G. Corb\`o and L. Maiani,
Nucl. Phys. B155 (1979) 93\semi
G. Corb\`o, Nucl. Phys. B212 (1983) 99\semi
G. Altarelli, N. Cabibbo, G. Corb\`o, L. Maiani and G. Martinelli,
Nucl. Phys. B208, (1982) 356\semi
A. Ali and E. Pietarinen, Nucl. Phys. B154 (1979) 519\semi
A. Czarnecki, M. Jezabek, and J.H. K\"uhn,
Phys. Rev. Lett. 73 (1994); Nucl. Phys. B351 (1991) 70\semi
A. Czarnecki and M. Jezabek, Nucl. Phys. B427 (1994) 3.}\nref\kuhn{M. Jezabek
and J.H. K\"uhn, Nucl.
Phys. B320 (1989) 20.}~\refs{\pwk,\kuhn}.
For decays involving a massive lepton in the final state,
parton model $\alpha_s$ corrections to the total decay rate
of $\ol B \to X_u \tau \ol \nu$ have been computed~\ref\ball{
E. Bagan et. al., preprint TUM-T31-67/94, MPI-PhT/94-49,
UAB-FT-347, and hep-ph/9409440.} for vanishing final quark mass
and unpolarized tau lepton.
Corrections to the differential decay rate ${d\Gamma \over d q^2 }$
for $B$ mesons going into polarized or unpolarized tau leptons
have also been computed, for massless
or massive final quarks, in the $\tau \ol \nu$ rest frame~\ref\cjz{
A. Czarnecki, M. Jezabek, and J.H. K\"uhn, preprint  TTP-94-26,
hep-ph 9411282.}.

In this paper we compute the matrix element of
time ordered weak currents in ``off-shell'' $b$ quark states
to order $\alpha_s$.  We apply this to double differential
rates for $B$ meson and $\Lambda_b$ baryon semi-inclusive
semileptonic decays involving either massive or massless final
quarks, and either polarized or unpolarized tau leptons.
The off-shell result for the
time ordered product is useful should
one ever wish to calculate the $\alpha_s$ corrections to the next to
leading terms in the ${\Lambda_{QCD}\over m_b}$ expansion.  We leave this
for future work.
By computing the absorbtive part of the on shell value,
we arrive at the parton model results for
decays involving a tau, in terms of a one dimensional integral over
the neutrino energy. This decay rate
differs  from that involving an electron due to the
large $\tau$ mass.
After a short review of the operator product expansion and the smearing method
in section 2, section 3 studies
the kinematics of B decays, section 4 gives our result for the B meson
triple diferential inclusive decay width,
section  5 studies the analytic structure of the form factors, section 6
deals with divergence cancellations, section 7 gives the formulas
for polarized $\Lambda_b$ decays, and section 8 states our conclusions.
Our results for $B$ decays are summarized, in a form appropriate for
inclusion in a Monte Carlo simulation, in Appendix E.
\newsec{The Smearing Method}
In this section we briefly review the procedure of
Chay, Georgi, and Grinstein~\cgg\
by which one arives at the ``smeared'' parton model as the leading term
in a ${\Lambda_{QCD}\over m_b}$ expansion.
The initial step in this procedure
is an operator product expansion of the time ordered product of
quark weak currents in terms of local operators in the heavy quark
effective theory.
\OMIT{ \footnote{$^1$}{In refs~\bksv , which do not consider
$\alpha_s$ corrections,
an operator product expansion
is first constructed in the full QCD theory, rather than the heavy
quark effective theory.
The advantage of this is
that certain QCD matrix elements, such as
$\bra{B,v}\bar b \gamma^{\mu} b\ket{B,v}$ are known exactly.
Using the conventional normalization of heavy quark states,
the latter matrix element is just $v^{\mu}$.
However as soon as one
includes QCD loop effects this version
of the operator product expansion breaks down.  Beyond tree level,
the expansion in residual
momentum $k$ is only sensible in the heavy quark effective theory.}}
\eqn\ope{\int d^4x\exp\lbrace {i\qx} \rbrace
T\left( J^{\mu}(x)J^{\nu^{\dagger}}(0)\right)
= \sum_i C_i(q,v)\hat O_{i,v}(0).}
The local operators $\hat O_{i,v}(x)$ are made up of light degrees of
freedom (gluons and light quarks) and
the  $b$ heavy quark operator $h_b^v(x)$,  related to
the full QCD field by
\eqn\match{h_b^v(x)={1+\vslash \over 2} \exp(im_b \vx )b(x),}
with $v$  the four velocity of the $B$ meson.
The coefficient
functions $C_i(q,v)$ can then be obtained to finite order
in $\alpha_s$ by considering the
perturbative matrix elements
of both sides of equation~\ope\
between  (unphysical) free quark and gluon
states, where the $b$ quark states carry momentum
$m_bv+k$.  The quark mass used here is the mass parameter of the
heavy quark effective lagrangian.
All the dependence on the residual momentum $k$ is
contained in the matrix elements of local
operators containing covariant derivatives of the heavy quark field.
The resulting operator product expansion is thus an expansion
in off-shell momentum $k$.

After the expansion has been computed, it is reinserted
between $B$ meson states. One might hope
to calculate or at least paramaterize
the hadronic tensor in this way.  However the operator product expansion
breaks down near the perturbative threshold.  Since the
perturbative analytic structure of $T^{\mu\nu}$ is contained in the
coefficient functions,  the coefficient functions must be singular near
thresholds.  To see this explicitly, consider
the tree level expectation value of the current-current
correlator in a $b$ quark:
\eqn\cor{\bar{u}
{\gamma^{\mu}P_L(m_b\vslash-\qslash+\kslash+m_j)\gamma^{\nu}P_L  \over
(m_bv-q+k)^2-m_j^2+i\epsilon}u.}

The tree level coefficient functions for two quark operators are then
obtained by expanding the above expression as a Taylor series in $k$.
The tree level operator product expansion has been considered previously
by several authors~\bksv\ and the reader is referred to them
for details.
The important point is that the expansion parameter is
\eqn\exparam{{k\over (m_b v-q)^2-m_j^2+i\epsilon}.}
The denominator vanishes at threshold,  so the operator
product
expansion breaks down in its vicinity.
The discontinuity across the real axis in the complex $\vq$ plane
consists of an infinite series of derivatives of delta functions which appears
to be useless
at finite order:
\eqn\dlta{T_i=\sum_n c_n\delta^{(n)}\left( (m_b v-q)^2-m_j^2\right),}
where the $c_n$ are matrix elements of the general form
\eqn\cn{c_n=\bra{B}\bar {h_b^v}D^n h_b^v \ket{B}.}
The perturbative coefficient functions possess
singularities in the wrong places:
to arbitrary perturbative order in
$\alpha_s$ the threshold singularity is that of the parton model of
free quarks and gluons,
$(m_bv-q)^2=m_j^2$, rather than a real hadronic threshold.

These troubles may be avoided by computing $T_i(\vq,q^2)$
well away from
the physical region.
A large imaginary piece to $\vq$ acts as an infrared
cutoff, allowing one to compute the Wilson coefficients
perturbatively~\pqw.  Furthermore, the coefficients get
progressively smaller for higher orders in the
residual momentum $k$.   Far from the physical region, one expects
both the operator product expansion and QCD pertubation theory to
be well behaved.  Thus one can use
them to compute
integrals of $W_i(\vq,q^2)$ over real $\vq$.
Due to equation~\WT, such an integral
is equivalent to an integral  of $T_i(\vq,q^2)$ just above the cut
minus an integral just below.  The integration contour can
be deformed away from the real axis between the endpoints
of the integration. At a distance $m_b$ from threshold in
the complex $\vq$ plane the effective expansion parameter becomes
$k\over m_b$.
As long as these endpoints are far away
from both perturbative and real thresholds,  the error due
to the integration near the endpoints should be small.

\topinsert
\centerline{
\epsfysize=6cm
\epsfbox{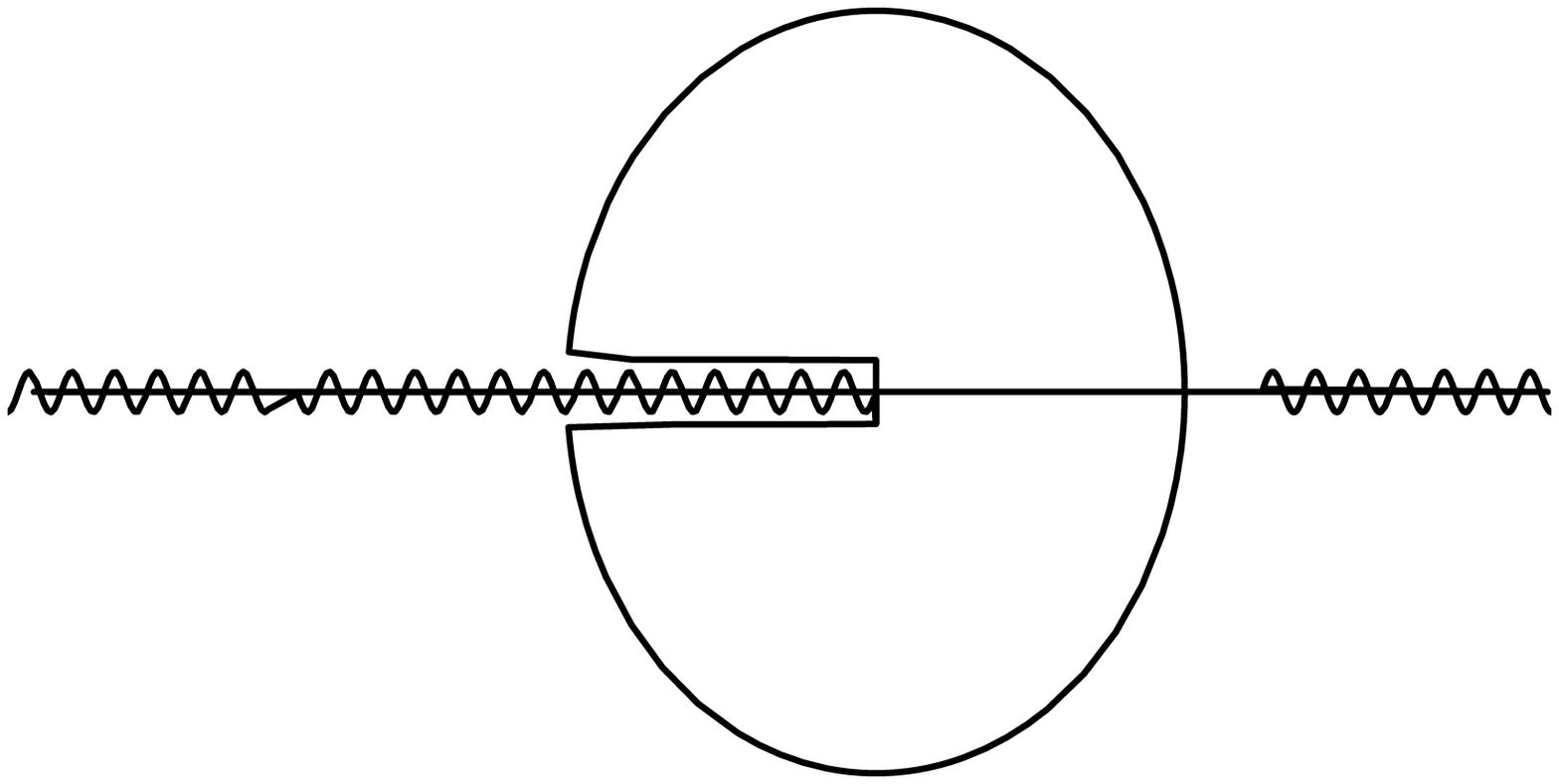}}
\vskip-2cm
\caption{{\bf Fig.~1}\hskip.3cm{Analytic structure of $T^{\mu\nu}$
and integration contour in the complex $\vq$ plane}}
\endinsert

In this paper we compute the differential decay width
${d^3\Gamma\over dE_{\tau}dq^2dE_{\bar \nu}}$. These three kinematic
variables may be varied as long as we treat the final state quark as
off-shell.  Pertubation theory does not give a sensible result for
this rate, but one can obtain sensible results
for ${d^2\Gamma\over dE_{\tau}dq^2}$, provided
that $E_{\tau}$ and $q^2$ are such that
the boundaries of the $\vq$ integration are not near threshold.
Integration over $\vq=E_{\tau}+E_{\bar\nu}$
is performed by integrating
the triple differential decay width
over neutrino energies at fixed tau energy.

In the parton model, the cut in the complex
$\vq$ plane is given by the B decay threshold
\eqn\thresh{\vq \le {m_b^2+q^2-m_j^2\over2m_b}.}
There exists another cut in
$\vq$, coming from the threshold for weak processes involving
two $b$ quarks in the final state.  This threshold is given
in the parton model by
\eqn\twob{\vq \ge {(2m_b+m_j)^2-m_b^2-q^2\over 2m_b}.}
For the particlar case
$m_j=0$ and $q^2=m_b^2$, this threshold coincides with the
 $B$ decay threshold \thresh.
In this case, near that point
 one can not get the operator product expansion to work
by integrating past the $B$ decay threshold, because there is no
way to close the integration contour in the complex plane
far from the region where the perturbative expansion fails.
The cuts together with the integration contour are drawn in fig.~1.

We shall compute the $B$ decay rate
using only the lowest order term in the operator
product expansion.  This calculation is equivalent to the parton
model calculation of the spin averaged decay rate of a free $b$ quark,
where the quark mass is equal to the mass parameter of the heavy quark
effective theory.
The reason for the equivalence is as follows.
The leading term of the operator product expansion
is determined from the expectation value
of the current correlator in an on shell $b$ quark.  This expectation
value is of the form $C_j(\vq,q^2,\alpha_s)
\bar u(m_bv) \Gamma_j u(m_bv)$,
where $\Gamma_j$ is some combination of Dirac matrices and $u$ is an
on shell spinor.
The corresponding term in the operator product expansion is
$C_j(\vq,q^2,\alpha_s)\bar {h_b^v}\Gamma_j h_b^v$
\footnote{$^2$}{Operators which contain gluons
or light quark pairs are sub-leading.  Thus, weak annihilation of the $b$
quark and light antiquark is suppressed in the $m_b\rightarrow\infty$
limit.}.
 This last statement is true to all orders in $\alpha_s$ due to
the heavy quark
spin symmetry:  at leading order in ${\Lambda_{QCD}\over m_b}$
any operator of the form $\bar {h_b^v}\Gamma_j h_b^v$ is a component of
a conserved current associated with heavy quark number or heavy quark
spin.
Furthermore the expectation value of such an operator in a $B$
meson is exactly known by the very same symmetry arguments; the
Isgur-Wise function at zero recoil is one.  Thus the
$B$ meson expectation value averaged over polarizations is equal
to the $b$ quark expectation value averaged over spins:
\eqn\quiv{T^{\mu\nu}=\frac12 C_i(\vq,q^2)Tr\left({1+\vslash\over
2}\Gamma_i\right),}
where we have normalized our heavy quark states to $v^0$.

\newsec{Kinematics}

It is convenient to decompose the hadronic tensor $T_{\mu\nu}$
into form factors,
\eqn\form{\eqalign{T^{\mu\nu} = -& g^{\mu\nu}T_1+v^{\mu}v^{\nu}T_2
                   -i\epsilon^{\mu\nu\alpha\beta}v_{\alpha}q_{\beta}
                   T_3\cr
                   +& q^{\mu}q^{\nu}T_4+
                   (q^{\mu}v^{\nu}+q^{\nu}v^{\mu})T_5\cr
                   +&(q^{\mu}v^{\nu}-q^{\nu}v^{\mu})T_6.\cr}}
$T_6$ vanishes due to the time reversal invariance of the strong
interactions: $T^{\mu\nu}(v,q)=T_{\nu\mu}(v^{\prime}q^{\prime})$
where $v^{\prime\mu}=v_{\mu}$ and $q^{\prime\mu}= q_{\mu}$.
$W_{\mu\nu}$
 may be similarly decomposed. The
decay rate  for spin averaged  $B$'s
is given by
\eqn\rate{\eqalign{
{d^3\Gamma \over dq^2 dE_\tau dE_{\nu} }
       &= {\abs{V_{jb}}^2\, G_F^2\over 2\pi^3}
\Bigl[W_1\(q^2-\mts\)+W_2\(2\en\et
        -{q^2-\mts \over 2}\) \cr
       &\hskip.7in+W_3\((\et-\en)(q^2-\mts)-2\en\mts\) \cr
       &\hskip.7in+\mts \(W_4({q^2-\mts \over 2})+W_5 (2\en)\)\ \ \Bigr].}}
For $\pm$ helicity tau's we have:
\eqn\pmrate{\eqalign{
{d^3\Gamma_\pm \over dq^2 dE_\tau dE_{\nu} }
       &=\frac12 {d^3\Gamma \over dq^2 dE_\tau dE_{\nu} }
\(1 \pm \frac\et\kleng\)\cr
      &\pm {\abs{V_{jb}}^2\, G_F^2\over 4\pi^3} \mts \frac\et\kleng
          \Bigl[-2W_1-W_2-2W_3\(\kleng \cos\alpha+\en\) \cr
       &\hskip1.7in+W_4\(\mts+2\et(\kleng \cos\alpha-\et)\)\cr
       &\hskip1.7in+W_5\(2(\kleng \cos\alpha-\et)\)\ \ \Bigr],}}
where the angle between the tau and neutrino trajectories is given by
\eqn\des{
\cos\alpha={\mts-q^2+2\en\et \over 2\en \kleng}.}
Note that this expression reduces to eq. \rate\  when summed over tau
helicities.
The leptonic variables $E_{\tau}$,  $q^2$ and $E_\nu=\vq -E_\tau$ are
constrained to the region given by
\eqn\limEtau{
m_\tau \le E_\tau \le {m_B \over 2} (1-{m_X^2-m_\tau^2 \over m_B^2}),}
\eqn\limqsquared{
m_\tau^2 \le \q^2 \le
{(E_\tau + \sqrt{E_\tau^2-m_\tau^2}) (m_B^2-m_X^2-2 E_\tau m_B)
+m_\tau^2 m_B \over m_B-E_\tau -
 \sqrt{E_\tau^2-m_\tau^2}},}
\eqn\limvq{E_{\tau}+{q^2-m_{\tau}^2\over 2(E_{\tau}
+\sqrt{E_{\tau}^2-m_{\tau}^2})}
\le
\vq\le {\rm Min}[E_{\tau}+{q^2-m_{\tau}^2\over 2(E_{\tau}
-\sqrt{E_{\tau}^2-m_{\tau}^2})},{m_B^2+q^2-m_X^2 \over 2 m_B}].}

$m_B$ is the B meson mass, and $m_X$ the mass of the lightest hadron
containing the final state $u$ or $c$ quark . With the usual assumption
$m_B > m_b$ , the allowed region of integration
for $\vq$ includes the parton model theshold \thresh.
The upper limit for $\vq$ shows a
new feature of tau kinematics; for a massless final lepton the upper
limit is simply the threshold value \thresh. The lower and upper bounds
in \limvq\ correspond to parallel and anti-parallel $\tau$ and $\nu_\tau$
trajectories.

In the case of a decaying polarized $\Lambda_b$ we can parameterize
the spin-dependent part of $T^{\mu \nu }$ in terms
 of nine additional form factors,
\eqn\Sdefined{\eqalign{
T^{\mu\nu}_S &= - \qs \Bigl[-g^{\mu\nu} G_1 + v^{\mu}v^{\nu} G_2 - i
\epsilon^{\mu\nu\alpha\beta} v_\alpha q_\beta G_3 + q^\mu q^\nu G_4 \cr
&\qquad +\left(q^\mu v^\nu + q^\nu v^\mu\right) G_5\Bigr] +
\left(s^\mu v^\nu + s^\nu v^\mu\right) G_6 +
\left(s^\mu q^\nu + s^\nu q^\mu\right) G_7 \cr
&\qquad
+i \epsilon^{\mu\nu\alpha\beta}v_\alpha s_\beta G_8
+i \epsilon^{\mu\nu\alpha\beta}q_\alpha s_\beta G_9,
}}
where $s$ is the spin vector of the $\Lambda_b$ baryon in the   $\Lambda_b$
rest frame.
Note that there are no terms of the form $i(q^\mu
\epsilon^{\nu\alpha\beta\lambda} v_\alpha q_\beta s_\lambda
-  (\mu \rightarrow \nu))$ or $ i(v^\mu \epsilon^{\nu\alpha\beta\lambda}
v_\alpha q_\beta s_\lambda - (\mu \rightarrow \nu)).$ They can be shown to
vanish using the identity
$$      g^{\mu\nu} \epsilon^{\alpha\beta\lambda\sigma}
-g^{\mu\alpha} \epsilon^{\nu\beta\lambda\sigma} + g^{\mu\beta}
\epsilon^{\alpha\nu\lambda\sigma} -g^{\mu\lambda}
\epsilon^{\nu\alpha\beta\sigma} +g^{\mu\sigma}
\epsilon^{\nu\alpha\beta\lambda}= 0. $$
The differential decay rate in this case has additional terms that depend on
the spin of the $\Lambda_b$:
\eqn\spinrate{\eqalign{
&{d^4\Gamma_\pm \over dq^2 dE_\tau dE_{\nu} d\!\cos\theta} =\
        \frac12\ {d^3\Gamma_\pm \over dq^2 dE_\tau dE_\nu } \cr
        &+\cos\theta\Bigl[\
         \frac12\ {d^3\Gamma_\pm^{\(W_i \rightarrow G_i \)} \over dq^2 dE_\tau
dE_\nu } \ \
        (\en \cos\alpha +\kleng)\ \cr
        &\hskip.5in+ {\abs{V_{jb}}^2\, G_F^2\over 8\pi^3}\Bigl[\(1 \pm
\frac\et\kleng\)\Bigl(\ G_6\(-2\en(\et \cos\alpha)+\kleng\)\cr
        &\hskip1.5in-G_7\(2\mts\en \cos\alpha\)-G_8\(2\en (\et \cos\alpha
-\kleng)\) \cr
        &\hskip1.5in-G_9 \((q^2 -\mts)(\en \cos\alpha -\kleng)+2\mts\en
\cos\alpha\)\Bigr)\cr
        &\hskip1.35in\pm \mts \frac{2\en}\kleng \Bigl(
G_6\(\cos\alpha\)+G_7\(\et \cos\alpha -\kleng\) \cr
        &\hskip1.5in\hskip.98in +  G_8\(\cos\alpha\)
+G_9\(\cos\alpha(\en+\et)\)\Bigr)\Bigr] \Bigr],}}
where $\theta$ is the angle between the spin vector of the $\Lambda_b$ and the
$\tau$ three-momentum, and where
in the second term
 $d^3\Gamma_\pm^{\(W_i \rightarrow G_i\)}$ is identical to $d^3\Gamma_\pm$,
except that all the $W_i$'s are replaced by the spin-dependent $G_i$'s.

Note that when we integrate the polarized cross section \spinrate\
 over $\cos \theta$ the term proportional to $\cos \theta$ vanishes,
while the constant term gets multiplied by a factor
of $2$ and so reproduces eq.~\pmrate .
\newsec{B meson decay}
To compute the time ordered product of eq.~\WT, we must evaluate
the graphs in figs.~2,3,4 and 5. We do this by
first separating the Dirac structure from the loop integrals,
evaluating the loop integrals in terms of a function we call
S, and reducing the Dirac algebra with the help of the heavy quark effective
theory.
We present analytic formulas for S and
express the form factors $T_i$ in terms of S and its derivatives.
After that, it is a matter of numerical integration to
generate plots at arbitrary values of $q^2$ and
$E_\tau$, with kinematic cuts, or
with additional refinements. Such refinements might include corrections
due to the W propagator~\kuhn, or $m_j/m_b$ corrections arising from
expressing the width in terms of quark masses evaluated at one scale
instead of on-shell \ref\nir{Y. Nir, Phys. Lett. B 221, (1989) 184.}.
In addition, higher order corrections
due to these graphs (such as the $\alpha_s$ correction to the
$\frac1{m_b^2}$ operator $\bra{B} (\im D)^2 \ket{B}$) may be computed by
substituting $p \to m_b v + k$ and expanding in $\frac{k}m_b$. Strong
corrections to decays of polarized $\Lambda_b's$ may also be found
straightforwardly by evaluating the Dirac traces between spinors instead
of averaging over spin.

We do our calculation in minimal subtraction and Feynman gauge,
with a gluon mass $\lambda$ to regulate
infrared divergences.  For a massless final state, there is considerable
simplification if one uses dimensional regularization instead of
a gluon mass \ref\pecci{B. Guberina, R.D. Peccei, and R. R\"uckl,
Nucl. Phys. B171 (1980) 333.}, but
for $B \to c l \bar \nu$, the simplification is less obvious.
For the charm case, the amplitudes may be expressed in terms of
generalized hypergeometic functions, which can be numerically integrated
over the charm momentum and analytically integrated over the lepton
momenta \ref\pham{Q.Ho-kim and X. Y. Pham, Ann. Phys. 155 (1984) 202.}.
However, if lepton spectra are desired,
this trick for analytic
integration over lepton momenta is not useful.

We devide our calculation into several parts.  The total
form factor is expressed as the sum of the contributions
of various diagrams
$$T_i = T_i^0 + T_i^{self} + T_i^{vert} + T_i^{box}.$$

\OMIT{Strong corrections to
$B \to X_q e \nu$ have been computed previously using a gluon
mass\ref\corbo{G. Corb\`o, Nucl. Phys. B212 (1983) 99}.}

\subsec{Heavy Quark Self Energy corrections}
   The quark matrix element of the tree level graph of fig.~2 is
\eqn\treegraph{
   \ub {\ga^\mu (\pslash -\qslash) \ga^\nu P_L\over
       [(p-q)^2 - m_j^2 + \im \epsilon]} \u, }
where $p$ is the momentum of the external b quark, $q$ is the W
momentum, and $\u$ is the b quark spinor. For the parton model
result, we will eventually take $p = m_b v$, with $v^2 =1$.

\topinsert
\centerline{
\epsfysize=4.5cm
  \epsfbox{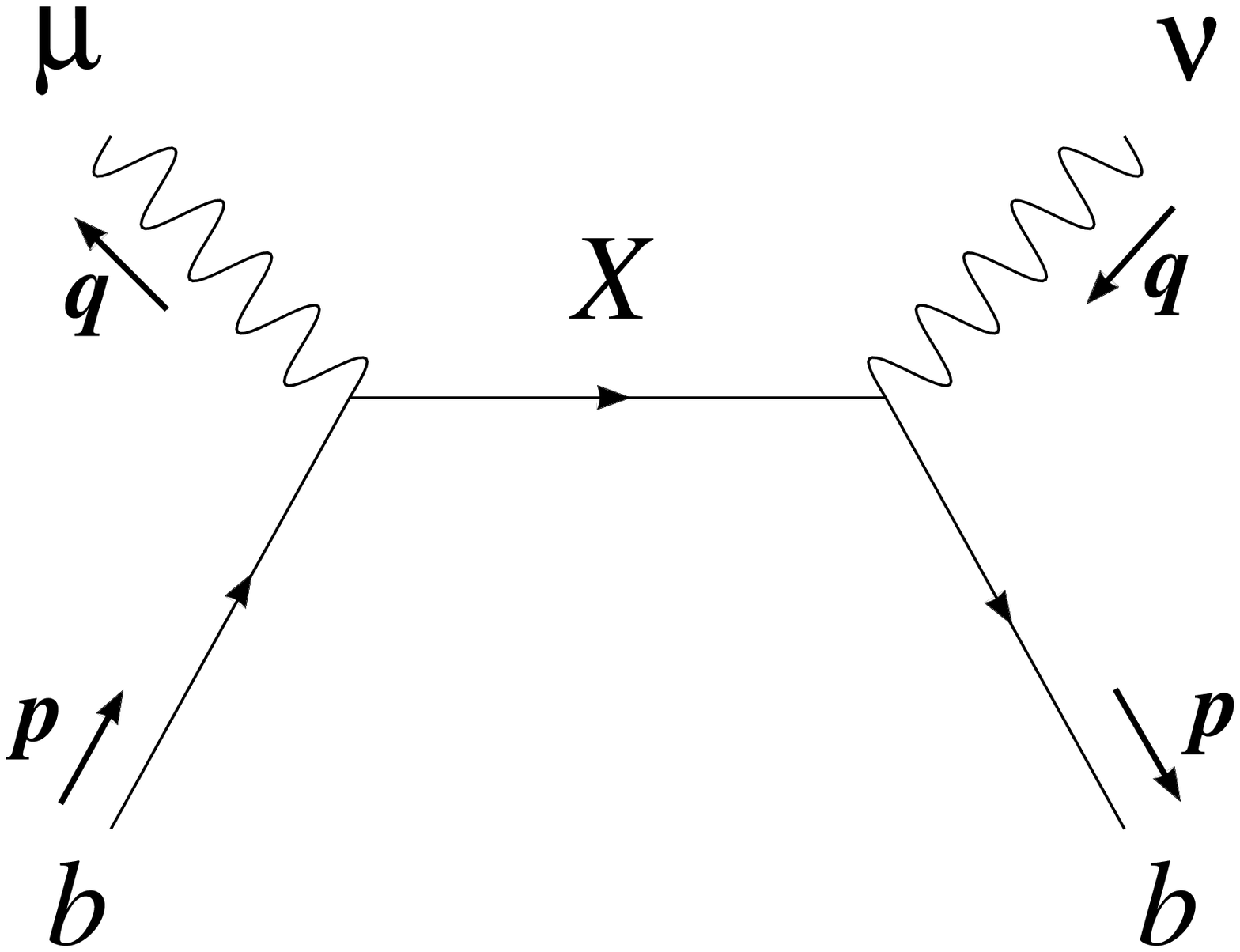}}
\caption{{\bf Fig.~2}\hskip.3cm{Tree level contribution to inclusive
semi-leptonic $B$ decay.}}
\endinsert

Wave-function renormalization of the external heavy quark lines
modify the tree level diagram by the multiplicative factor
$$ T^{\mu\nu}_{tree} \to T^{\mu\nu}_{tree} (1+\delta Z_b),$$
where $\delta Z_b = \pd{\pslash} \Sigma |_{\pslash=m_b}$, and
$- \im \Sigma$ is the self-energy loop of a $b$ quark with momentum
$p$.
This loop gives, for $D= 4 - \epsilon$ dimensions,
$$- \im \Sigma = - {\im \alpha_s \over 3 \pi}
   [ (\epsilon-2) \pslash B_{00}(p,0,m_b) + (4-\epsilon)
     m_b B_0(p,0,m_b) ] ,$$
where
\eqn\bo{\eqalign{
B_0(p,0,m_b) &= \frac{-\im}{\pi^2} \int {\tpid d^Dl \over \gprop
    \bprop} \cr
    &= \frac2\epsilon + \ln (4 \pi) -\gamma
     +2 + \im \pi {p^2 - m_b^2 \over p^2} \cr
       &- \frac{m_b^2}{p^2} \ln ({m_b^2 \over \mu^2})
       - {p^2 - m_b^2 \over p^2} \ln ({p^2 - m_b^2 \over \mu^2})\cr}}
and
\eqn\boo{\eqalign{
p^\al B_{00}(p,0,m_b) &=
\frac{-\im}{\pi^2} \int {\tpid d^Dl \quad l^\al \over \gprop \bprop} \cr
    &=p^\al [ -\frac1\epsilon -\frac12 \ln 4\pi +\frac12\gamma
               -1 + \im \pi {m_b^4 - p^4 \over 2 p^4}\cr
                &+ \frac{m_b^4}{2 p^4} \ln \frac{m_b^2}{\mu^2}
           - \frac{m_b^2}{2 p^2} + \frac{p^4 - m_b^4}{2 p^4}
              \ln\frac{p^2 - m_b^2}{\mu^2}].\cr }}

We may set the gluon mass to zero for the self-energy, which is
infrared finite, but for the derivative of the self-energy the
gluon mass $\lambda$ is needed to
regulate the infrared divergence. In this case, we get
$$ \dz_b=\pd{\pslash} \Sigma |_{\pslash=m_b} = \alpi
\[-\frac2\epsilon-\ln 4\pi
+\gamma-4+\ln\frac{m_b^2}{\mu^2}-2\ln\frac{\lambda^2}{m_b^2}\].$$

The resulting form factors are
\eqn\trez{\eqalign{
  T_1^{0} &= \frac1{2 \D}(m_b-\vq)(1 + \dz_b)\cr
  T_2^{0} &= \frac1{\D} m_b ((1 + \dz_b)\cr
  T_3^{0} &= \frac1{2 \D} ((1 + \dz_b)\cr
  T_4^{0} &= 0 \cr
  T_5^{0} &=  -\frac1{2 \D} ((1 + \dz_b),\cr }}
where $\D = (p-q)^2 -m_j^2 + \im \epsilon$.
\subsec{Final Quark Self Energy}
Since we are using the optical theorem, the light c or u quark
self-energy is not incorporated into wave-function renormalization, but
instead arises as a separate graph.  These contributions to the
form factors enter in the  same proportions as the
tree result, since self-energy corrections don't alter the
Lorentz structure of the leading order operator.
    The self-energy graph of fig.~3, which also accounts for final
state radiation, gives

\topinsert
\centerline{
\epsfysize=4.5cm
\epsfbox{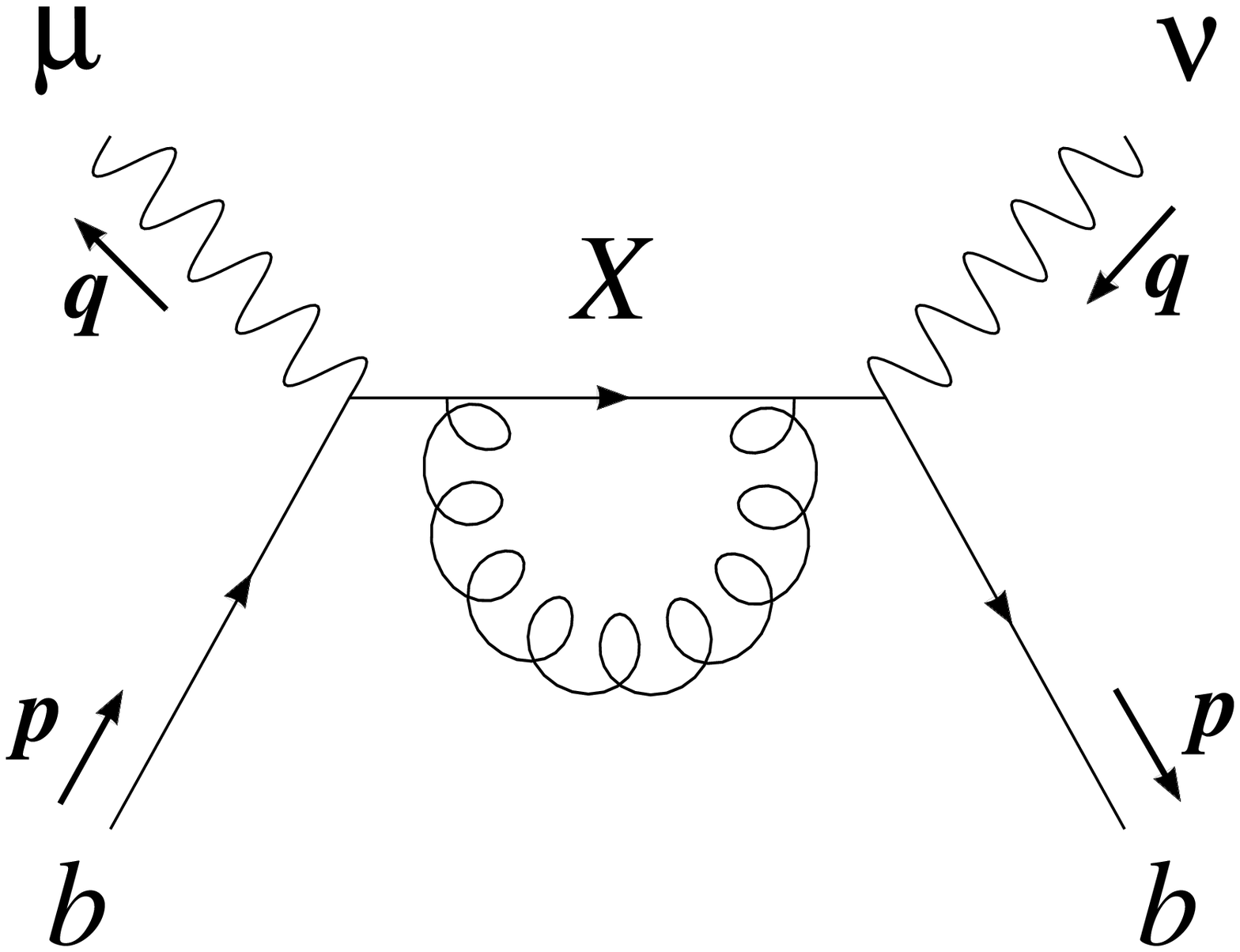}}
\caption{{\bf Fig.~3}\hskip.3cm{Final quark self-energy
 diagram.}}
\endinsert

$$\eqalign{
   {\im \alpha_s \over 3 \pi}{\ub \ga^\mu (\pslash -\qslash) \ga^\nu P_L \u
\over (\D)^2 } & \bigl[  2 m_j^2 (4-\epsilon)
B_0(p-q,0,m_j)+
 \cr & ((p-q)^2 + m_j^2)(\epsilon-2) B_{00}(p-q,0,m_j)\bigr].}$$
It gives a form factor contribution of
\eqn\Tselfs{\eqalign{
T_1^{self} =&\,{\alpha_s\over 3 \pi}\,{1\over 2\Delta_0^2}\,
(m_b-\vq \,)\,
        \CA[(p-q)^2,m_j^2]\cr
\noalign{\smallskip}
T_2^{self} =&\,{\alpha_s\over 3 \pi}\,{1\over \Delta_0^2}\,m_b\,
         \CA[(p-q)^2,m_j^2]\cr
\noalign{\smallskip}
T_3^{self} =&\,{\alpha_s\over 3 \pi}\,{1\over 2\Delta_0^2}\,
          \CA[(p-q)^2,m_j^2]\cr
\noalign{\smallskip}
T_4^{self} =&\,0\cr
\noalign{\smallskip}
T_5^{self} =&\,{\alpha_s\over 3 \pi}\,{-1\over 2\Delta_0^2}\,
         \CA[(p-q)^2,m_j^2],          }}
where
\eqn\defA{\eqalign{
\CA[(p-q)^2,m_j^2]&=\left({2\over \eps} -\gamma
+\ln(4\pi\mu^2)\right)(-(p-q)^2 + 7 m_j^2)
         -(p-q)^2 +10 m_j^2-{m_j^4\over (p-q)^2}\cr &
   -\left( {m_j^6\over (p-q)^4}-7{m_j^4\over (p-q)^2}-(p-q)^2+7m_j^2
\right)
          \ln((m_j+\lambda)^2-(p-q)^2-i
\epsilon)\cr &
   +\left( {m_j^6\over (p-q)^4}-7{m_j^4\over (p-q)^2}\right)
            \ln(m_j^2).\cr }}

The optical theorem tells us that to obtain the B decay rate
we need to compute the imaginary part of the $T_i$'s. In the case of
the heavy quark self-energy, as $\dz_b$ is real, there is only one
contribution, coming from
\eqn\imdelta{Im{\lbrace {1\over(x+i\eps)}\rbrace}=-\pi \delta(x).}

On the other hand, $Im(T^{self})$ has two contributions, a first one
from
\eqn\imdeltaprime{Im{\lbrace {1\over{(x+i\eps)^2}}\rbrace}=\pi
\delta^\prime(x),}
and a second one from the imaginary part of the function
$\CA[(p-q)^2,m_j^2]$.
  Those two contributions correspond to two
different cuts in the Feynman diagram. In the first, we cut
the final quark propagator near the weak vertex and put it
on-shell with the Dirac delta. The virtual gluon renormalizes the
light quark wave-function. It is easy to see, by integrating the
$\delta$ and $\delta^\prime$ functions over $\vq$, that
the  resulting initial and final quark wave-function
renormalizations are identical,
after the obvious interchange $m_b \leftrightarrow m_j$.
The second contribution corresponds to cutting through the loop. In this
case, the gluon is real, corresponding to final state bremsstrahlung.
The fact that $\CA$ only has a non-zero imaginary part
when $(p-q)^2 \geq (m_j+\lambda)^2$ immediately implies the
parton model branch cut
\thresh. $\CA$ has an additional dependence on the gluon mass which cancels
the infrared divergences that arise from the vertex corrections.
The analogous infrared divergence from bremsstrahlung off the initial quark is
contained in the box diagram, fig.~5.


\subsec{Vertex Corrections}
The virtual gluon correction to the weak vertex is
\eqn\vv{
-\frac{\im 4 g^2}3 \lint {\ga^\al(\lslash+m_b)\ga^\mu P_L(\lslash
-\qslash +m_j) \ga_\al \over \gprop \cprop\bprop}. }
This enters into both vertex graphs in fig.~4, giving

\topinsert
\centerline{
\epsfysize=4cm
\epsfbox{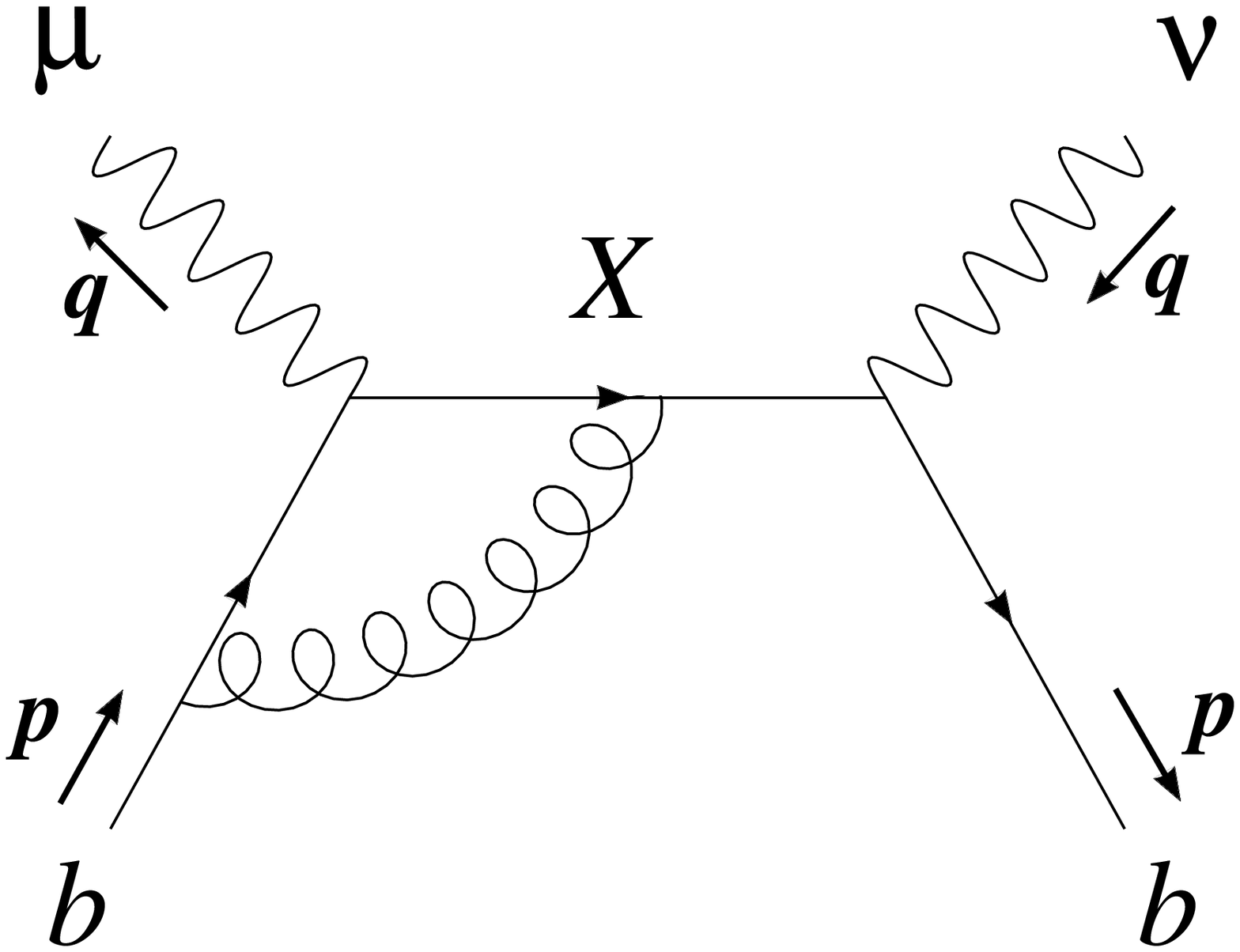}
\hskip 1.5cm
\epsfysize=4cm
\epsfbox{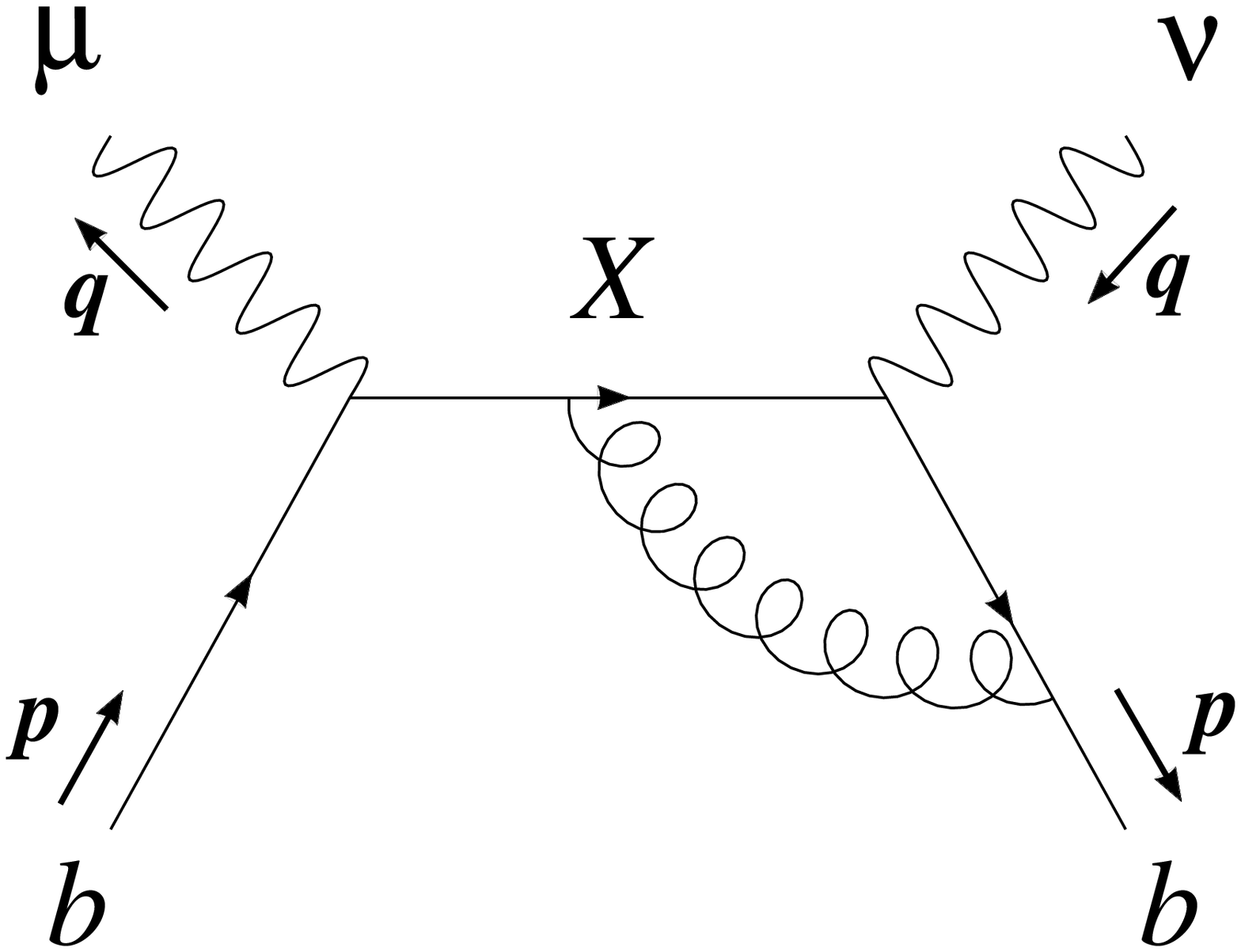}}
\caption{{\bf Fig.~4}\hskip.3cm{Vertex correction diagrams.}}
\endinsert

\eqn\vertgraph{
 {\im \alpha_s \over 3 \pi \D }
\ub [\NV C + \NV_\alpha C^\alpha + \NV_{\alpha\beta} C^{\alpha\beta}] \u , }
where
\eqn\Cdef{
  C^{[.,\alpha,\alpha\beta]} =
  \frac{-\im}{\pi^2} \int {\tpid d^Dl\quad [1, l^\alpha, l^\al l^\be] \over
  \bprop \cprop \gprop} }
and
\eqn\NVdef{\eqalign{
  \NV &= m_b \ga^\la \ga^\mu (\ga_\la m_j^2 - \qslash \ga_\la \pslash')
       \ga^\nu P_L +m_b \ga^\mu (\ga^\la m_j^2 - \pslash' \ga^\la \qslash)
         \ga^\nu P_L \ga_\la \cr
  \NV_\al &= \ga^\la\ga_\al\ga^\mu(\ga_\la m_j^2 - \qslash \ga_\la \pslash')
   \ga^\nu P_L +m_b \ga^\la \ga^\mu\ga_\al\ga_\la\pslash' \ga^\nu P_L\cr
    &+ \ga^\mu (\ga^\la m_j^2 - \pslash' \ga^\la \qslash)\ga^\nu P_L \ga_\al
      \ga_\la +m_b \ga^\mu\pslash'\ga^\la \ga_\al \ga^\nu P_L\ga_\la \cr
  \NV_{\al\be} &= \ga^\la\ga_\al\ga^\mu \ga_\be \ga_\la \pslash'
      \ga^\nu P_L + \ga^\mu \pslash' \ga^\la\ga_\al\ga^\nu
      P_L \ga_\be \ga_\la, \cr
     }}
with $p'= p-q$.
The Dirac algebra from eq.~\NVdef\ is dramatically simplified by
using heavy quark symmetry.
Spinors $\u$ correspond to operators $b = h_b^{(v)} + \O(\frac1m_b)$, so
to leading order in \lom, we have
\eqn\trace{
 \bra{B} \ub \Gamma \u \ket{B} =
   \frac12 \Tr[\Gamma \frac{1 + \vslash}{2}] .}
The same formula applies to $B^*$ mesons if we average over
spin.  Higher order corrections in \lom\ can be computed by inserting
heavy quark symmetry breaking terms from the heavy quark Lagrangian
\ref\mw{ N.~Isgur and M.~B.~Wise, \pl{232}{1989}{113}; \pl
{237}{1990}{527}\semi
E.~Eichten and B.~Hill, \pl{234}{1990}{511}\semi
H.~Georgi, \pl{240}{1990}{447}.}.

The integrals $C^{[.,\alpha,\alpha\beta]}$ may be expressed, by means of
the usual Feynman parametrization, momentum shifting, and integration,
in terms of the Feynman parameter integral
\eqn\Sdef{
S(a,b,c,d,e,f)=\int_0^1 dx\int_0^{1-x} dy\ln X(x,y;{\bf a}) ,}
where
$$X(x,y;{\bf a})={1\over \mu^2} (a x + b y + c x^2 + d y^2 + e x y + f),$$
and
\eqn\inputs{\eqalign{
a &= -(p^2 - m_b^2 + \la^2)  \cr
b &= m_j^2- \la^2- (p-q)^2 \cr
c &= p^2 \cr
d &= (p-q)^2 \cr
e &= 2 p \cdot (p-q) \cr
f &= \la^2 - \im \epsilon .
  }}
The real and imaginary parts of this function are evaluated
analytically
in Appendices A and B. The special case of a massless final state quark,
which would give spurious divergences if
the zero mass were substituted in the formulas
in the previous appendices, is considered in Appendix C.
These closed expressions are exact, and have been verified numerically.
However, they are cumbersome to handle, and it may be
simpler to evaluate these functions numerically than analytically.
Expressions for the $C$ integrals in terms
of $S$ and its derivatives are given in appendix D.

After evaluating the traces and integrals in terms of S and
its derivatives, we arrive at the following results:
\eqn\Tvertexs{\eqalign{
T_1^{vert} =&\,{-\alpha_s\over 3 \pi}\,{2\over \Delta_0}\,\,[
      (-m_b+ \vq)\,({1\over \eps} - {\gamma \over 2}
       + \ll - 1 - \,S)\cr &+ m_b
       m_j^2 S_{,f} - m_b^3\,(2 S_{,a} +2 S_{,b} - S_{,c} - S_{,d} -2
S_{,e}
       - S_{,f})\cr &+
      \, m_b^2 \vq \,(3 S_{,a} +5 S_{,b} - S_{,c} -3 S_{,d} -4 S_{,e}
      -2 S_{,f}) \cr &+
      \, m_b q^2 \,( S_{,a} + S_{,d} - S_{,f})\,+\,q^2 \,\vq\,
         ( S_{,b} - S_{,d})\cr &-
      \,2 m_b (\vq)^2\,( S_{,a} +2 S_{,b} - S_{,d} - S_{,e} - S_{,f})
]
\equiv {\alpha_s \over 3 \pi \Delta_0} \hat T^{vert}_1
\cr
\noalign{\smallskip}
T_2^{vert} =&\,{\alpha_s\over 3 \pi}\,{4\over \Delta_0}\,\,[
         m_b\,({1\over \eps} - {\gamma \over 2} +\ll- 1 - \,S)\cr &+\,
    (m_b^3 + m_b m_j^2 - m_b q^2)\,( S_{,a} + S_{,b} - S_{,f})\,+ m_b q^2
S_{,e})
]
\equiv {\alpha_s \over 3 \pi \Delta_0} \hat T^{vert}_2
\cr
\noalign{\smallskip}
T_3^{vert} =&\,{\alpha_s\over 3 \pi}\,{2\over \Delta_0}\,\,[
         ({1\over \eps} - {\gamma \over 2}+\ll -1 - \,S)\cr &+\,
    m_b^2 \,(3 S_{,a} +3 S_{,b} - S_{,c} - S_{,d} -2 S_{,e}- 2 S_{,f})\cr
&+
      \,q^2\,( S_{,b} - S_{,d})\,-\,2 m_b \vq\,
      ( S_{,a} +2 S_{,b} - S_{,d} - S_{,e}- S_{,f}) ]
\equiv {\alpha_s \over 3 \pi \Delta_0} \hat T^{vert}_3
 \cr
\noalign{\smallskip}
T_4^{vert} =&\,{-\alpha_s\over 3 \pi}\,{4\over
\Delta_0}\,m_b\,(S_{,a}-S_{,e})
\equiv {\alpha_s \over 3 \pi \Delta_0} \hat T^{vert}_4
\cr
\noalign{\smallskip}
T_5^{vert} =&\,{-\alpha_s\over 3 \pi}\,{2\over \Delta_0}\,\,[
         ({1\over \eps} - {\gamma \over 2} +\ll-1 - \,S)\cr &+\,
    m_b^2 \,( S_{,a} +2 S_{,b} -2 S_{,f})\,+ m_j^2 S_{,b}\cr &-\,2\,
    m_b \vq\,( S_{,a} + S_{,b}- S_{,e} - S_{,f}) ]
\equiv {\alpha_s \over 3 \pi \Delta_0} \hat T^{vert}_5
.\cr           }}

The notation $S_{,a}$ means the derivative of $S(a,b,c,d,e,f)$ with
respect to $a$.
Again, we have two contributions to $Im(T^{vert})$: one from the
imaginary part of the light quark propagator and one from the
imaginary part of $S$ and its derivatives. The former corresponds
to the virtual gluon corrections to the vertex, and the latter to
bremsstrahlung due to interference between initial and
final state radiation.
\subsec{Box Diagram}
The box graph, depicted in fig.~5, gives

\topinsert
\centerline{
\epsfysize=5.5cm
\epsfbox{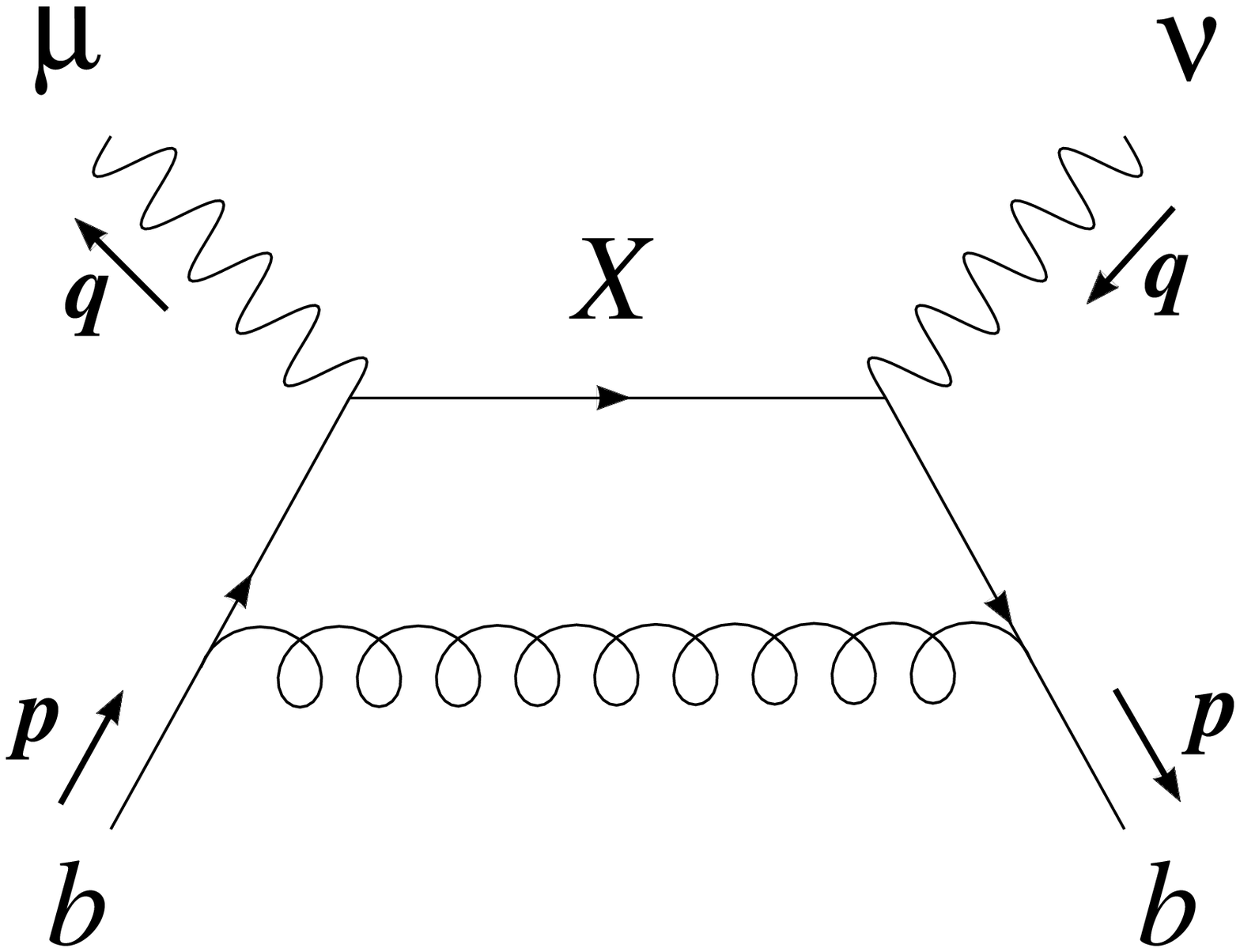}}
\caption{{\bf Fig.~5}\hskip.3cm{Box diagram.}}
\endinsert

\eqn\boxgraph{\eqalign{
\frac{4 g^2}3\ub &\lint
{\ga^\al(\lslash+m_b)\ga^\mu P_L(\lslash -\qslash) \ga^\nu
 (\lslash+m_b)\ga_\alpha \over \gprop\cprop\bprop^2}\u \cr
&=\frac{\im \alpha}{3\pi} \[ \NB D + \NB_\alpha D^\alpha
+ \NB_{\alpha\beta} D^{\alpha\beta}
    +  \NB_{\alpha\beta\gamma} D^{\alpha\beta\gamma}\]\u,\cr
     }}
where
\eqn\Ddef{
  D^{[.,\alpha,\alpha\beta,\alpha\beta\gamma]} =
  \frac{-\im}{\pi^2} \int {\tpid d^Dl\quad
 [1, l^\alpha, l^\al l^\be, l^\al l^\be l^\ga] \over
  \bprop^2 \cprop \gprop} }
and
\eqn\NBdef{\eqalign{
  \NB &= -m_b^2 \ga^\la \ga^\mu P_L\qslash \ga^\nu \ga_\la \cr
\NB_\al &=m_b^2 \ga^\la \ga^\mu P_L\ga_\al\ga^\nu\ga_\la \cr
      &- m_b \ga^\la \ga_\al \ga^\mu P_L\qslash \ga^\nu \ga_\la
     - m_b \ga^\la\ga^\mu P_L\qslash \ga^\nu\ga_\al\ga_\la\cr
\NB_{\al\be} &=m_b  \ga^\la \ga_\al \ga^\mu P_L \ga_\be \ga^\nu \ga_\la \cr
&+ m_b \ga^\la \ga^\mu P_L\ga_\al\ga^\nu\ga_\be\ga_\la
- \ga^\la \ga_\al \ga^\mu P_L\qslash \ga^\nu \ga_\be\ga_\la \cr
\NB_{\al\be\ga} &= \ga^\la \ga_\al \ga^\mu P_L
    \ga_\be \ga^\nu \ga_\ga \ga_\la. \cr
    }}
Evaluating the traces and integrals as with the vertex diagrams yields
\eqn\Tboxs{\eqalign{
T_1^{box} =&\,{-\alpha_s\over 3 \pi}\,[m_b (3 S_{,a}+S_{,c}+S_{,e}) +
              m_b^3(-4 S_{,a,a} -4 S_{,a,b} + S_{,a,c} \cr
         &\,+ S_{,a,d} +2 S_{,a,e}
                 +2 S_{,a,f} + S_{,c,c} +3 S_{,c,d} +3 S_{,c,e} +
S_{,d,e})\cr
    &\,+m_b q^2(-2 S_{,a,b} + S_{,a,d} -2 S_{,a,e} + S_{,c,d} + S_{,d,e})
   - \vq\,\, (S_{,a}+ S_{,e})\cr
  &\,+m_b^2 \vq\,\,(2 S_{,a,a} +6 S_{,a,b} + S_{,a,c} - S_{,a,d} -2
S_{,a,f}
       -6 S_{,c,d} -3 S_{,c,e} -3 S_{,d,e})\cr
     &\,+q^2 \vq\,(S_{,a,d} - S_{,d,e})
        + m_b (\vq)^2(-2 S_{,a,d} +2 S_{,c,d} +2 S_{,d,e})]
\equiv{\alpha_s \over 3 \pi} \hat T^{box}_1 \cr
\noalign{\smallskip}
T_2^{box} =&\,{\alpha_s\over 3 \pi}\,\,2\,[-m_b ( S_{,a} - S_{,c} - S_{,e})
\cr &\,+
             m_b^3(4 S_{,a,a} +4 S_{,a,b} - S_{,a,c} - S_{,a,d} -2
S_{,a,e}
                 -2 S_{,a,f} - S_{,c,c} \cr
                &\,-3 S_{,c,d} -3 S_{,c,e} - S_{,d,e})
      -m_b q^2( S_{,a,d} - S_{,c,d} - S_{,d,e})]
\equiv{\alpha_s \over 3 \pi} \hat T^{box}_1 \cr
\noalign{\smallskip}
T_3^{box} =&\,{\alpha_s\over 3 \pi}\,\,[ -S_{,a} +3 S_{,e} \cr
           &\,-m_b^2(-2 S_{,a,a} -2 S_{,a,b} - S_{,a,c}
         + S_{,a,d} +2 S_{,a,f} -2 S_{,c,d} - S_{,c,e} - S_{,d,e})\cr
                &\,-q^2( S_{,a,d} - S_{,d.e})
                - m_b \vq\,(-2 S_{,a,d} +2 S_{,c,d} +2 S_{,d,e})]
\equiv{\alpha_s \over 3 \pi} \hat T^{box}_1 \cr
\noalign{\smallskip}
T_4^{box} =&\,{-\alpha_s\over 3 \pi}\,\,4\,[-m_b ( S_{,a,b} + S_{,a,e} -
S_{,c,d} - S_
{,d,e})
                   + \vq\,( S_{,a,d} - S_{,d,e})]
\equiv{\alpha_s \over 3 \pi} \hat T^{box}_1 \cr
\noalign{\smallskip}
T_5^{box} =&\,{-\alpha_s\over 3 \pi}\,\,[- S_{,a} + S_{,e} \cr
           &\,-m_b^2(-2 S_{,a,a} -6 S_{,a,b} - S_{,a,c}
         + S_{,a,d} +2 S_{,a,f} +6 S_{,c,d} +3 S_{,c,e} +3 S_{,d,e})\cr
                &\,-q^2( S_{,a,d} - S_{,d.e})
                - m_b \vq\,(2 S_{,a,d} -2 S_{,c,d} -2 S_{,d,e})]
\equiv{\alpha_s \over 3 \pi} \hat T^{box}_1 .\cr
}}

The function $S_{,a,b}$ denotes the second derivative of $S(a,b,c,d,e,f)$ with
respect to $a$ and $b$.
There is only one way to obtain an imaginary part from the $T^{box}$,
via the imaginary part of $S$ and its derivatives. This corresponds to
cutting through the loop to obtain the
 bremsstrahlung
due to initial state radiation.

\newsec{Analytic Structure of Form Factors}

The form factors $W_i(\vq,q^2)$ are given by the discontinuity of
$T_i(\vq,q^2)$ across the cut in the complex $\vq$ plane.
The general location of the physical cuts
may be deduced from the expression~\tord\
by inserting a complete sets of states.
There is a  cut along the real axis
for
\eqn\cutone{\vq \le {m_B^2+q^2-m_X^2 \over 2m_B}.}
This cut corresponds to weak amplitudes for
which the initial hadronic state is a $B$ (containing one $b$ quark
and one light antiquark), and the final state contains one j type quark
($u$ or $c$)
and the same light antiquark.
$m_X$ is the mass of the lightest such hadronic final state.
The weak decay $B\rightarrow X\tau\bar\nu_{\tau}$ corresponds to the
portion of this cut for which $\vq>\sqrt{q^2}$
(since physical final state lepton pairs have
${|\vec q|}^2 \ge 0$).  By crossing symmetry
one obtains  cuts in other portions of the above region,
corresponding to different physical processes, such as $B \nu_\tau
\rightarrow X \tau$.

There is one other class of cuts arising from the
$x^0<0$ portion of the integral in equation~\tord, and
corresponding to the production of two $b$ quarks in
the final state.  This cut lies
along the real axis for
\eqn\cuttwo{\vq \ge {m_X^{\prime 2}-m_B^2-q^2\over 2m_B},}
where $m_X^{\prime}$ is the mass of the lightest
hadronic state containing two b quarks  which can be
produced in a semileptonic weak process involving a $B$
in the initial state.  This process occurs only at order $\alpha_s^2$.

\OMIT{
and the corresponding cut is situated at the right of the order
$\alpha_s$ one.
In this region the discontinuity of $T^{\mu\nu}$
across the cut does not correspond to the
$W^{\mu\nu}$ defined in equation~\had .}

   The location of the branch cuts for the form factors $T_i$ in the complex
$\vq$ plane for
fixed $q^2$ can be explicitely determined by looking at the discontinuities
of the $T$'s along the real $\vq$
axis. For the vertex and box diagrams, this
means looking for an imaginary part  of $S(\ba)$. $Im\[S(\ba)\]$  is
simply the area within the triangular integration region over which the
argument $X(x,y;\ba)$ is negative, the branch cut is determined by
studying the curve $X(x,y;\ba)=0$.  After expressing $\ba$ in terms of
physical variables, it is easy to show that $Im\[S(\ba)\] \ne 0$ if and
only if $ (p -q)^2 > m_j^2$. This implies a branch cut for
$\vq < {m_b^2 + q^2 - m_j^2 \over 2 m_b}$, in agreement with ref~\cgg .
For the final quark self-energy, we have to look for an imaginary part
in $\CA\[(p-q)^2,m_j^2\]$, which leads us to the same cut.

  An alternate way to see this is to consider two particle unitarity.
Unitarity of the amplitude $A_{if}$ between initial state $\ket{i}$
and final state $\bra{f}$ implies
$$A_{fi} - A_{if}^* = \sum_n \im (2 \pi)^4 \delta^{(4)}(p_i -p_n)
                          A_{fn}A_{in}^*.$$
For a two particle intermediate state (in our case a gluon and a quark),
the right hand side becomes
$$\frac{i}2 {k \over 16 \pi^2 E_T} \int d\Omega
      \bra{f}A\ket{k_1,k_2} \bra{k_1,k_2}A^\dagger\ket{i}
     \theta(E_T^2 - m_j^2),$$
where $k$ is the magnitude of the three-momentum of either intermediate
particle in the center of mass frame, and $E_T$ is total center of mass
energy. Since
$${k\over E_T}={1\over 2} \sqrt{(1 - \frac{m_j^2}{(p-q)^2})^2},$$
two particle unitarity provides a positive check on our calculation.
\newsec{Divergence Cancellation}
\subsec{Ultraviolet Divergences}
We performed the loop momentum integration in $4-\epsilon$ dimensions in
order to regulate the ultraviolet divergences.
The divergent part of each graph is always proportional
to the factor
\eqn\div{{2 \over \epsilon}-\gamma+\ln(4\pi\mu^2).}
The final and initial quark self-energies contribute the same
ultraviolet divergence, while the box is convergent.
One can see from \trez\ and \Tvertexs\ that the divergent
contribution from the vertex is -2 times that of the initial
quark self-energy, so the dependence on the regulator $\epsilon$
and the renormalization scale $\mu$ cancel exactly. This is nothing
but an explicit application of the Ward identity to our case.

\subsec{Infrared Divergences}
The cancellation of infrared divergences (which we have regulated
with a gluon mass) is more involved. The triple differential
decay distribution is IR divergent; the dependence on the gluon mass
cancels only after integration over $\vq$.
The self-energies exhibit the IR divergence explicitely as $\ln \lambda^2$
terms, but in the vertex and box diagrams the divergences are hidden in the
$S$ functions. It can be seen by numerical evaluation that of all
the $S$ functions, only $S_{,f}$ and $S_{,a,f}$ blow up for zero gluon mass.
These two functions can be
written as a gluon independent term plus another term proportional to
$\ln \lambda^2$. Upon integration over $\vq$, we have verified that the
coefficients of these logarithms cancel
exactly the IR logarithms from the self-energy. This non trivial
cancellation of the gluon mass dependence give us confidence
in our formulas. We have numerically verified this cancellation for
up and charm final quark masses and for several values of $q^2$. The
cancellation is independent of the tau energy, since all the $T$
form factors are independent of  $E_\tau$.

\OMIT{
We can again cut our diagrams to understand phisically each
contribution. The pieces were we exchange the imaginary part of
$\Delta_0$'s in the denominator for $\delta(\Delta_0)$ or
$\delta^\prime(\Delta_0)$ represent the effect of virtual gluons.
We have the self-energy loops for both quarks and the vertex correction
which are IR divergent. In this last case the divergence is hidden
in the real part of $S_{,f}$ at theshold. When we take the imaginary part
from somewhere else (the function $\CA$ for the final quark self-energy,
$S_{,f}$ for the vertex and $S_{,a,f}$ for the box), we have the
different contributions to the bremsstrahlung, the effect of real gluons.
As always, the gluon mass dependence between real gluon and virtual
gluon contributions exactly cancel each other and leave a finite
result.}
\newsec{$\Lambda_b$ baryon decay}
We now  calculate the additional spin-dependent form factors $G_1$
through $G_9$ that describe the decay of a polarized $\Lambda_b$.
The Dirac algebra is a little more complicated in this case since we
cannot use equation eq.~\quiv.
First we reduce the expressions given in eqs.~\NVdef\ and \NBdef\ by means
of the Chisholm identity
\eqn\chisholm{
\gamma^\mu \gamma^\alpha \gamma^\nu = g^{\mu\alpha} \gamma^\nu +
g^{\nu\alpha} \gamma^\mu - g^{\mu\nu} \gamma^\alpha + i
\epsilon^{\mu\nu\alpha\beta}\gamma_\beta \gamma_5.}
\OMIT{and inserting $\vslash $'s which are equal to the identity when
acting on the heavy $b$-quark spinors.i}
 Note that in our convention
$\epsilon_{0123}=+1$. After some  algebra which we performed with the help of
Feyncalc~\ref\feyncalc{R. Mertig, Guide to FeynCalc 1.0,Universit\"at
W\"urzburg, February 1992.}\ the Dirac structure
of the N's reduces to terms with zero or one gamma matrix and $\gamma_5$'s.
 The matrix elements of these operators can be further simplified
using the heavy quark symmetry. To first order in $1/m_b$ we have
\eqn\matrixelem{\eqalign{
\bra{\Lambda_b (v,s) } \bar h_b \Gamma h_b \ket{\Lambda_b (v,s) }&= \bar u(v,s)
\Gamma u(v,s),\cr
 \bar u(v,s) u(v,s) &= 1,\cr
 \bar u(v,s) \gamma_5 u(v,s) &=0,\cr
 \bar u(v,s) \gamma^\lambda u(v,s) &=v^\lambda,\cr
 \bar u(v,s) \gamma^\lambda \gamma_5 u(v,s) &=s^\lambda.\cr}}
We now contract the resulting expressions with the $C$'s given in Appendix
D and collect terms with respect to their Lorentz structure. The baryon form
factors $T_1$ to $T_5$, which enter into the first line of equation
\spinrate, are identical to the meson form factors $T_1$ to $T_5$
in equations \trez, \Tselfs, \Tvertexs, and \Tboxs. The spin-dependent
form factors in the rest of \spinrate\ are found by comparing the
above result to the defining equation for the $G$'s, eq. \Sdefined.

We divide the spin dependent form factors into three pieces
$$G_i = G_i^0 + G_i^{vert} + G_i^{box}.$$
The results from wavefunction renormalization and the self-energy graph
are proportional to the tree level result, so they may be combined
into
\eqn\selfSs{\eqalign{
G_1^{0} &= {-1 \over 2 \Delta_0} (1 + \dz_b
            + {\alpha_s\over 3 \pi\Delta_0}  \CA[(p-q)^2,m_j^2])\cr
G_6^{0} &= {-m_b \over 2 \Delta_0} (1 + \dz_b
               + {\alpha_s\over 3 \pi\Delta_0}  \CA[(p-q)^2,m_j^2])\cr
G_7^{0} &= {1 \over 2 \Delta_0} (1 + \dz_b
               + {\alpha_s\over 3 \pi\Delta_0}  \CA[(p-q)^2,m_j^2])\cr
G_8^{0} &= {m_b \over 2 \Delta_0} (1 + \dz_b
              + {\alpha_s\over 3 \pi\Delta_0}  \CA[(p-q)^2,m_j^2])\cr
G_9^{0} &= {-1 \over 2 \Delta_0} (1 + \dz_b
              + {\alpha_s\over 3 \pi\Delta_0}  \CA[(p-q)^2,m_j^2])\cr
G_2^{0} &= G_3^{0}=  G_4^{0}= G_5^{0}=0, \cr }}
where the functions $D^0$ and $\CA$ are defined in eq.s \trez\ and \defA.

The contributions to the $G$'s from the vertex diagrams are
\eqn\vertSs{\eqalign{
G^{vert}_1=&\tscoef {2\over\Delta_0}(-(\ovrep)+1+S-m_b^2(3\sa+3\sb-\sc-\sd\cr
     &-2\se-2\sf)-q^2(\sb-\sd)+2m_b\vq(\sa+2\sb-\sd-\se-\sf))\cr
G^{vert}_2=&\tscoef {4m_b^2\over\Delta_0}(-\sa+\sc+\se)\cr
G^{vert}_3=&\tscoef {m_b\over\Delta_0}\se\cr
G^{vert}_4=& 0\cr
G^{vert}_5=&\tscoef {2m_b\over\Delta_0}(\sa-\se)\cr
G^{vert}_6=&\tscoef {2m_b\over\Delta_0}(-(\ovrep)+1+S-m_j^2(\sa+\sb-\sf)\cr
     &-m_b^2(\sa+\sb-\sf)+q^2(\sa+\sb-\se-\sf)\cr
     &-m_b\vq(\sa-\sc-\se))\cr
G^{vert}_7=&\tscoef
{2\over\Delta_0}(\ovrep-1-S+m_j^2\sb-m_b\vq(\sa+2\sb-\se-2\sf)\cr
    &+m_b^2(2\sa+2\sb-\sc-\se-2\sf))\cr
G^{vert}_8=&\tscoef
{m_b\over\Delta_0}(2(\ovrep)-2-2S-q^2(2\sa+2\sd+\se-2\sf)\cr
     &-2m_j^2\sf+2m_b^2(2\sa+2\sb-\sc-\sd-2\se-\sf)\cr
     &-m_b\vq(3\sa+6\sb-\sc-4\sd-5\se-2\sf))\cr
G^{vert}_9=&\tscoef {1\over\Delta_0}(-2(\ovrep)+2+2S-2q^2(\sb-\sd)\cr
     &-m_b^2(3\sa+4\sb-\sc-2\sd-3\se-2\sf)\cr
     &+m_b\vq(4\sa+8\sb-4\sd-3\se-4\sf)).\cr}}

The contribution to the $G$'s from the box are calculated in the same
way as the vertex contributions. Again, the algebra is a little
tedious but straightforward. Making repeated use of the Chisholm
identity eq.~\chisholm\ , the heavy quark matrix elements eq.~\matrixelem ,
 and contracting with the $D$'s eq.~\Ddef\  we arrive at
\eqn\theboxSs{\eqalign{
G^{box}_1=&\tscoef (\sa +\se - q^2 (\sad -\sde) + 2m_b\vq (\sad-\scd -\sde)\cr
     &- m_b^2 ( 2\saa + 2\sab -\sac+\sad-2\saf-2\scd-\sce-\sde))\cr
G^{box}_2=&\tscoef 4 m_b^2 (\sab-2\sad-2\sae+2\scd+\sce+\sde)\cr
G^{box}_3=&\tscoef m_b(-\sab+\sad-\sae)\cr
G^{box}_4=&\tscoef 4 (-\sad+\sde)\cr
G^{box}_5=&\tscoef 2m_b(-\sab+3\sad+\sae-2\scd-2\sde)\cr
G^{box}_6=&\tscoef (m_b^3(-4\saa-4\sab+3\sac+3\sad+6\sae+2\saf-\scc\cr
     &-3\scd-3\sce-\sde)+m_b(\sa-\sc-\se)+m_b q^2 (\sad-\scd-\sde)\cr
     &+2 m_b^2 \vq (\sab-2\sad-2\sae+2\scd+\sce+\sde))\cr
G^{box}_7=&\tscoef (-\sa+\se +q^2(-\sad+\sde)+m_b^2(2\saa+4\sab\cr
     &-\sac-3\sad-4\sae-2\saf+2\scd+\sce+\sde)\cr
     &-2 m_b (\sab-2\sad-\sae+\scd+\sde))\cr
G^{box}_8=&\tscoef (m_b^3(4\saa+4\sab-3\sac-3\sad-6\sae-2\saf+\scc+3\scd\cr
     &+3\sce+\sde)+m_b q^2 (3\sab-4\sad-\sae+\scd+\sde)\cr
     &-3m_b(\sa-\sc-\se)- m_b^2 \vq (3\saa+7\sab-\sac-7\sad-8\sae\cr
     &-2\saf+4\scd+2\sce+2\sde))\cr
G^{box}_9=&\tscoef (\sa-3\se+q^2(\sad -\sde)-m_b\vq(\sab+\sad+\sae-2\scd\cr
     &-2\sde)+ m_b^2 (\saa +\sab -2\scd-\sce-\sde)).\cr}}

The self-energy diagram gives no spin dependent contribution.
\newsec{Conclusions}
We have presented $\alpha_s$ corrections to triple differential
inclusive distributions for $B \to X_q \tau \ol\nu$, in closed analytic
form.
We have also calculated the dependence of the distribution on the tau
polarization, for both up and charm quarks.
We have not performed the integration over neutrino energy which is
necessary to give the physically meaningful
double differential distribution, but have instead
left the answer in terms of a one dimensional integral which can
be evaluated numerically. An additional integration
over tau energy or momentum transfer will yield
${d\Gamma \over d q^2}$ or ${d\Gamma \over d E_\tau}$, respectively.
Formulas appropriate for insertion in a Monte Carlo program are
presented in Appendix E.

We have verified numerically the cancellation of infrared divergences
due to soft gluons, and the stability of the double differential
distribution against variations in the infared regulating gluon mass. In
addition, the
formulas for the graphs and integrals entering the triple differential
distribution have been checked extensively numerically.

Our computation has been in the context of a heavy quark operator
product expansion. An advantage of this is that extending the
calculation
to polarized $\Lambda's$ is simple. The quartic distribution for
polarized Lambda's to polarized tau's $+ \ol\nu + X_q$ involves
nine additional form factors, which we have computed. Again, integration
over the neutrino energy is necessary to get an experimentally
observable distribution.

The formulas in this paper may be useful for purely theoretical reasons,
in addition to experimental ones.  A computation of the $\alpha_s$
corrections to the Wilson coefficeints of the operator product expansion
may give insight on how perturbation theory breaks down as one nears
the endpoint of the lepton spectrum.  This requires calculating one-loop
graphs in QCD (which we have done) and comparing them to one-loop graphs
in the effective theory.

Strong corrections to form factors
at $\CO({1\over m_b^2})$ may also be of interest~\ref\lksv{
M. Luke and M.J. Savage, Phys.Lett.B321 88 (1994).}.
The structure of our results is amenable to such a calculation, by
expressing the $b$ momentum as $p= m_b v + k$ and expanding in residual
momentum $k$. The only complication here is that care must be taken
not to cut the b quark lines when computing Bremstrahlung; this requires
a modification of our formula for the imaginary part of $S$.

\bigskip
\noindent{\bf{Acknowledgements}}

We thank Aneesh Manohar for helpful suggestions and discussions.
We acknowledge the support of the Department of Energy, under contract
DOE--FG03--90ER40546.  One of us (FJV) is supported by the Ministerio de
Educaci\'on y Ciencia (Spain).

\bigskip
\noindent{\bf{Appendix A: Evaluation of Re(S) }}

  We want to evaluate
$$S(a,b,c,d,e,f)=\int_0^1 dx\int_0^{1-x} dy\ln X(x,y;{\bf a})$$
where
$$X(x,y;{\bf a})=\frac1\mu (a x + b y + c x^2 + d y^2 + e x y + f).$$
We evaluate the real and imaginary parts of S separately because
we will use a change of variables which causes
our integration contour to cross branch cuts for some values
of our parameters, producing spurious contributions of $\im \pi$.
In addition, this change of variables is valid
only if $d \ne 0$. For an on-shell up quark, we may either take the
limit as $d \to 0$, or use the formula in appendix C.

First, we perform a change of variables, $x \to 1 -x$
and $y \to y - \alpha x$, where
$$\alpha = \frac1{2 d} (e + \sqrt{e^2 - 4 c d}) ,$$
and use the identity
$$\int_0^1 dx \int_{-\alpha x}^{(1-\alpha)x} dy =
\int_0^{1 -\alpha} dy \int_{\frac{y}{(1-\alpha)}}^1 dx
- \int_0^{-\alpha} dy \int_{-\frac{y}{\alpha}}^1 dx.$$
In the first integral on the right hand side, make the replacement
$y \to (1 -\alpha) y$, and in the second, replace
$y \to -{y \over \alpha}$. The resulting equality is valid only for
the real parts:
$$Re[S(\ba)]= Re
\{\int_0^1dy\int_y^1dx\left[(1-\alpha)\ln X(\ba_1)+\alpha\ln X(\ba_2)\right]\}
  ,$$
where
$$\eqalign{
 \ba_i &= \{a_i, b_i, c_i, d_i, e_i, f_i\}, \cr
   a_1 &= \alpha (b + e) - a - 2 c = a_2, \cr
   b_1 &= (1 -\alpha)(b + e), \qquad b_2 = -\alpha (b + e), \cr
   c_1 &= 0 = c_2, \cr
   d_1 &= (1 - \alpha)^2 d, \qquad d_2 = \alpha^2 d , \cr
   e_1 &= (1 -\alpha)(2 d \alpha - e), \qquad e_2 = \alpha ( e - 2 d
\alpha), \cr
   f_1 &= f + a + c = f_2. \cr
           }\eqno(A.1)$$
This expression can be evaluated straightforwardly by completing the
squares inside each integral. The final result is
$${ Re[S(\ba)] =Re[ (1 - \alpha)F(\ba_1) + \alpha F(\ba_2)]}\eqno(A.2)$$
where
$$F(\ba)= \frac1e \[ -f(\ba_3) + f(\ba_4)\],$$

$$\eqalign{
    f(\ba) &= a \{ g(a,b,c,d) + g(a,-b,c,d) -1
     + 3 c - 2d \cr &+(\frac12 - c +d)
  \ln a + (c^2 - b^2)(1 - \ln a)\[\ln(c+d) - \ln(1+c+d)\]\},\cr}\eqno(A.3)$$

$$\eqalign{
    g(a,b,c,d)&= \ln(b+d)\[\frac12 b^2 + b c + c d -
     \frac12 d^2 + (b^2 -c^2)\ln({c+d \over c-b})\]\cr
    +\ln(1+b &+d)\[\frac12 - \frac12 b^2 -c - b c + d -c d + \frac12 d^2
    + (c^2 -b^2)\ln({1+c+d \over c-b})\] \cr
    &+(c^2 - b^2)\[Sp({1+b+d \over b-c}) - Sp({b +d \over b -c}) \]
           ,\cr}\eqno(A.4)$$

$$\eqalign{
     a_3 &= d+e,   \qquad    a_4 = d, \cr
     b_3 &= \sqrt{({a+b \over 2 d +2 e})^2 - \frac{f}{d+e}},
        \qquad b_4 =  \sqrt{({b+e \over 2 d})^2 - \frac{a+f}{d}},\cr
     c_3 &= \frac{a}e - {a+b \over 2d +2e}, \qquad c_4=
          \frac{a}e - {b+e \over 2d},\cr
     d_3 &= {a+b \over 2d +2e}, \qquad d_4= {b+e\over 2d}\cr
     e_3 &=e_4=f_3=f_4=0
      ,\cr}\eqno(A.5)$$
and the Spence function is defined so that
$$Sp(x) = \int_x^0 {\ln(1-t)\over t} dt.$$
Note that in the evaluation of, for example, $F(\ba_1)$, the
parameters which enter into the definitions of $\ba_3$ and $\ba_4$ in
equation (A.5) are $\ba_1.$

\bigskip
\noindent{\bf{Appendix B: Evaluation of Im(S) }}

It is much easier to evaluate
the imaginary part of S than S itself.
$$ Im S=  \int_0^1 dx \int_0^{1-x} dy \,\,Im \ln X(x,y;{\bf a}) $$
is just $\pi$ times  the area in which X is negative, inside the triangle
in the $(x,y)$ plane defined by the integration limits (recall that
$f = \la^2 -\im \epsilon$ is the only complex input parameter).
 To obtain this area,
we solve the equation $X=0$, which gives us a hyperbola in the $(x,y)$ plane,
and integrate x as a function of y (or viceversa) in the triangle.
The evaluation of this area gives:
$$ Im S= \pi(A(\sqrt{b^2-4\,d\,f})-A(-\sqrt{b^2-4\,d\,f})), $$
where the function A is
$$\eqalign{
A(K) = &
{{a\,\left( b - K \right) }\over {4\,c\,d}} -
  {{e\,{{\left( b - K \right) }^2}}\over {16\,c\,{d^2}}} +
  \left( {{2\,b\,c - a\,e}\over
          { - 4\,c\,({e^2-4\,c\,d})}} +
        {{-b + K}\over {8\,c\,d}} \right) \times       \cr  &
  \sqrt{{{2\,{a^2}\,{d^2} - 2\,a\,b\,d\,e +
             {b^2}\,{e^2} - 2\,d\,{e^2}\,f +
             2\,a\,d\,e\,K - b\,{e^2}\,K}\over {{2
d^2}}}}\, \cr +
                     &
 {\left(a\,b\,e - {b^2}\,c   -
        {a^2}\,d    - f({e^2-4\,c\,d}) \right)   \over
    {\left(  {e^2-4\,c\,d} \right) ^{3\over 2}}}
       \,\ln [-4\,b\,c + 2\,a\,e -
        {  - ({e^2-4\,c\,d})  \,
            \left( b - K \right) \over d} + \cr &
        \,{\sqrt{{e^2-4\,c\,d}}}\,
         {\sqrt{{{4\,{a^2}\,{d^2} + 2\,(2\,a\,d\,e -
                {b}\,{e^2})(K-b) - 4\,d\,{e^2}\,f
              }\over {{d^2}}}}
           }] .}\eqno(B.1)$$

All the derivatives of S needed for the decay rate can be obtained from this
formula.

\bigskip
\noindent{\bf{Appendix C:
 Evaluation of Re(S) at threshold, for massless final quarks}}

The formula for the real part of S breaks down for $d=0,$ and it is
cumbersome to take the limit numerically. It is convenient to
have available a simpler representation of S which is valid when $d=0$.
The evaluation is straightforward, so we omit details and present only
the final result:
$$\eqalign{
  Re[S(a,b,c,0,e,f)] &= Re [-\frac12 +\frac12\ln(c-e)\cr
     &-\frac{c}e \ln(c)[\frac12 + 2 a_1
     -\frac{b}e + (c_1^2 +(-a_1 + \frac{b}e)^2)\ln({b\over b+e})]\cr
     &+\frac1e \ln(c-e) [\frac{c}2 + a_2 c + t_3 + t_4\ln(
                              {1 + a_2 + t_2\over a_2 + t_2})]\cr
    &+\frac{c}e f_0(a_2,\frac{t_3}c,c_2)
  - \frac{c}e f_0(a_1,a_1 -\frac{b}e,c_1) - f_0(a_2,-1-a_2,c_2)\cr
    &+ t_1 g_0(a_1,-a_1 + \frac{b}e,c_1)
     + \frac{t_4}e g_0(a_2,t_2,c_2) ] \cr }\eqno(C.1)$$
where
$$f_0(a,b,c)= f_{00}(a,b,c) + f_{00}(a,b,-c) - \frac12
          -a - 2b,$$
$$g_0(a,b,c)= g_{00}(a,b,c) + g_{00}(a,b,-c),$$
$$\eqalign{
   f_{00}(a,b,c)&=
     (-\frac12 a^2 - a b - b c + \frac12 c^2)\ln(a+c)\cr
 &+ (\frac12 +a + \frac12 a^2 + b + a b + b c - \frac12 c^2)
     \ln(1+a+c),\cr }\eqno(C.2)$$
$$\eqalign{
   g_{00}(a,b,c)&= \ln({1+a+b\over b-c})\ln(1+a+c)
       -\ln({a+b\over b-c})\ln(a+c) \cr
       &+ Sp({1+a+c\over c-b}) - Sp({a+c\over c-b}), }\eqno(C.3)$$
and
$$\eqalign{
    a_1 &= {a \over 2 c} \cr
    a_2 &= {a-b+e\over 2 (c-e)} \cr
    c_1 &= \sqrt{\frac{a^2}{4 c^2}- \frac{f}c }\cr
    c_2 &= \sqrt{\frac{(a-b+e)^2}{4 (c-e)^2} - \frac{b+f}{c-e}  }\cr
    t_1 &= \frac{c}e [ c_1^2 - (\frac{b}e - a_1)^2]\cr
    t_2 &= \frac{b}e -a_2 \cr
    t_3 &= a - 2 c a_2 - c t_2 \cr
    t_4 &= c t_2^2 + (2 c a_2 -a) t_2 + f - a a_2 + c a_2^2. \cr
    }\eqno(C.4)$$

The real part is taken because, as in Appendix A, the changes of
variable contribute extraneous factors of $\im \pi$  to the imaginary
part of S. While both equations (A.2) and (C.1) are exact, single
precision numeric evaluations of them may work poorly when the intermediate
parameters are anomalously large or small (for example, when $c \approx
e$ above).
\bigskip
\noindent{\bf{Appendix D: The loop integrals in terms of S's}}

$$\eqalign{
C=&{ -S_{,f}}\cr\cr
C^{ \alpha}=&{ -S_{,b}}\,\,q^{ \alpha}+
  \left( { S_{,a}} + { S_{,b}} - { S_{,f}} \right)\,
   p^{\alpha}\cr \cr
C^{ \alpha \beta}=&(-\frac12 S\, +
    \frac12\,(\frac1\eps-{\gamma \over 2}+\ll)\,)\,g^{ \alpha \beta}  -
  { S_{,d}}\,\,q^{\alpha}\,q^{ \beta} \cr &-
  \left( { S_{,b}} - { S_{,d}} - { S_{,e}} \right)\,
   \left( \, q^{ \beta}\,p^{\alpha} +
               q^{\alpha}\,p^{ \beta} \, \right) \cr &+
  \left( 2\,{ S_{,a}} + 2\,{ S_{,b}} - { S_{,c}} -
     { S_{,d}} - 2\,{ S_{,e}} - { S_{,f}} \right)\,
   p^{\alpha}\,p^{ \beta}  \cr \cr } \eqno(D.1)$$
$$\eqalign{
D=&{ -S_{,a,f}}\cr\cr
D^\alpha=& { -S_{,a,b}}\,\,q^{\alpha} +
  \left( { S_{,a,a}} + { S_{,a,b}} - { S_{,a,f}} \right)\,
   p^{\alpha}\cr\cr
D^{ \it \alpha \beta}=&-\frac12\, { S_{,a}}\, g^{ \alpha \beta}
  { -S_{,a,d}}\,q^{\alpha}\,q^{\beta} \cr &-
  \left( { S_{,a,b}} - { S_{,a,d}} - { S_{,a,e}} \right)\,
   \left( \, q^{\beta}\,p^{\alpha} +
               q^{\alpha}\,p^{\beta} \, \right) \cr &+
  \left( 2\,{ S_{,a,a}} + 2\,{ S_{,a,b}} - { S_{,a,c}} -
     { S_{,a,d}} - 2\,{ S_{,a,e}} - { S_{,a,f}} \right)\,
   p^{\alpha}\,p^{\beta}\cr\cr
D^{ \it \alpha \beta \gamma}=&-\frac12\, { S_{,e}}\,
\left(\,g^{ \it \beta \gamma}\,q^{\alpha}
       + g^{ \it \alpha \gamma}\,q^{\beta}
       + g^{ \it \alpha \beta}\,q^{\gamma} \,\right) -
  { S_{,d,e}}\,\,q^{\alpha}\,q^{\beta}\,q^{\gamma}\cr &-
  \frac12\,\left( { S_{,a}} - { S_{,c}} - { S_{,e}} \right)\,
      \left(\,g^{ \alpha \gamma}\,p^{\beta} +
                 g^{ \beta \gamma}\,p^{\alpha} +
                 g^{ \alpha \beta}\,p^{\gamma}\,\right) \cr &-
  \left( { S_{,a,d}} - { S_{,c,d}} - { S_{,d,e}} \right)\,
   \left(\,q^{\alpha}\,q^{\gamma}\,p^{\beta} +
        q^{\beta}\,q^{\gamma}\,p^{\alpha} +
        q^{\alpha}\,q^{\beta}\,p^{\gamma}\,\right) \cr &-
  \left( { S_{,a,b}} - 2\,{ S_{,a,d}} - 2\,{ S_{,a,e}} +
     2\,{ S_{,c,d}} + { S_{,c,e}} + { S_{,d,e}} \right)\,\cr &
   \left( \,q^{\alpha}\,p^{\beta}\,p^{\gamma}  +
       q^{\beta}\,p^{\alpha}\,p^{\gamma}  +
       q^{\gamma}\,p^{\alpha}\,p^{\beta} \,\right) \cr &+
  ( 3\,{ S_{,a,a}} + 3\,{ S_{,a,b}} - 3\,{ S_{,a,c}} -
     3\,{ S_{,a,d}} - 6\,{ S_{,a,e}} \cr & \qquad\qquad - { S_{,a,f}} + {
S_{,c,c}} \,\,
     + 3\,{ S_{,c,d}} + 3\,{ S_{,c,e}} + { S_{,d,e}} ) \,
   p^{\alpha}\,p^{\beta}\,p^{\gamma}\ . \cr\cr} \eqno(D.2)$$

\bigskip
\noindent
{\bf{Appendix E: Double differential B meson Decay rate}}

We collect here the complete formula for the $\alpha_s$
correction to the double
differential B meson decay rate. It is expressed as a one dimensional integral
amenable to insertion into a numeric integration program incorporating
any relevant cuts and detector efficiencies.
The masses $m_\tau,m_b, m_j$ correspond to the tau lepton,
the $b$ quark, and the $u$ or $c$ quark masses, respectively.
$E_\tau$ and $E_\nu$ are the tau and neutrino energies,
$v$ is the velocity of the $B$ meson, and
$\vq = E_\tau + E_\nu$ in the center of mass frame.

The double differential rate is
$$\eqalign{&
 {d^2\Gamma \over dq^2 dE_\tau }
      = {-\alpha_s\abs{V_{jb}}^2\, G_F^2\over 6\pi^3}\Biggl[
\(q^2-\mts\) \Bigl(
(m_b-\vq_{th})\dcero + \hat T^{vert}_1|_{th}
\cr &- \int d \,\vq \bigl(
{(m_b-\vq)\ms \over 2 \dt^2}  + {Im(\hat T^{vert}_1)/\pi \over
\dt}+ {Im(\hat T^{box}_1)\over \pi} \bigr)\Bigr)
\cr &
+\(2\et (\vq|_{th}-\et)  -{q^2-\mts \over 2}\) \(
2 m_b\dcero + \hat T^{vert}_2|_{th} \)
\cr &- \int d \,\vq
\(2\et (\vq-\et)  -{q^2-\mts \over 2}\) \Bigl(
{m_b\,\ms\over   \dt^2}  \cr &+ {Im(\hat T^{vert}_2)/\pi \over
\dt}+
  {Im(\hat T^{box}_2)\over \pi} \Bigr)\cr &
+\( (2\et-\vq|_{th})(q^2-\mts)-2\mts (\vq|_{th}-\et) \)
\( \dcero + \hat T^{vert}_3|_{th}\)
\cr &- \int d \,\vq
 \( (2\et-\vq      )(q^2-\mts)-2\mts (\vq-\et      ) \)
\bigl( {\ms \over 2\dt^2}
\cr &+  {Im(\hat T^{vert}_3)/\pi \over
\dt}+
  {Im(\hat T^{box}_3)\over \pi} \bigr)
\cr &+ \({q^2-\mts \over 2}\)\mts
\Bigl( \hat T^{vert}_4|_{th} -
 \int d \,\vq \bigl(
 {Im(\hat T^{vert}_4)/\pi \over
\dt}+ {Im(\hat T^{box}_4)\over \pi}\bigr)\Bigr)
\cr &
+2\mts \Bigl(  (\vq|_{th}-\en)
(-\dcero + \hat T^{vert}_5|_{th})
\cr &- \int d \,\vq \(\vq-\en\) \bigl(
{-\ms\over 2\dt^2}
\cr &+{Im(\hat T^{vert}_5)/\pi \over
\dt}+ {Im(\hat T^{box}_5)\over \pi}\bigr)\Bigr)
  \Biggr],}\eqno(E.1)$$
where
$$\ms=
  {m_j^6\over (m_b^2+q^2-2m_b \vq)^2}-7{m_j^4\over
 (m_b^2+q^2-2 m_b \vq)}-m_b^2-q^2+2 m_b \vq+7m_j^2.$$
The $\hat T|_{th}$ functions are the real parts of
the $\hat T$ functions defined in \Tvertexs\ and \Tboxs,
evaluated at the parton model theshold $\vq|_{th}=(m_b^2+q^2-m_j^2)/2 m_b$.
The real and imaginary parts of $\hat T$ may be
expressed in terms of the real part of the function
$S$ evaluated in Appendices A and C, and the imaginary
part of $S$ evaluated in Appendix B.
Integration limits are given in (3.5)-(3.7). Note that the $\hat T$
functions
have  an imaginary part only for
$\vq \le (m_b^2+q^2-(m_j+\lambda)^2)/(2 m_b)$;
 this $\lambda$ dependence of the integration limits
cancels all gluon mass dependence upon integration over $\vq$,
in the limit $\lambda \rightarrow 0$.

\listrefs
\bye